\documentclass[12pt, preprint]{aastex}
\newcommand\cm{{\;\rm cm}}
\newcommand\pcc{{\;\rm cm}^{-3}}
\newcommand\Kel{{\;\rm K}}

\newcommand\yr{{\;\rm yr}}

\newcommand\Msun{{\;\rm\,M_\odot}}

\newcommand\kms{{\;\rm km\,s^{-1}}}

\newcommand\tcro{t_{\rm cross} }
\newcommand\pc{{\;\rm\,pc}}
\newcommand\kpc{{\;\rm kpc}}
\newcommand\Kcm{{\;\rm K\,cm^{-3}}}
\newcommand\simgt{\lower.5ex\hbox{$\; \buildrel > \over \sim \;$}}
\newcommand\simlt{\lower.5ex\hbox{$\; \buildrel < \over \sim \;$}}
\newcommand\xhat{\hat{\mathbf{x}} }
\newcommand\yhat{\hat{\mathbf{y}} }

\newcommand\vel{\mathbf{v}}
\newcommand\ergs{{\rm \;erg\,s^{-1}}}

\shorttitle{SPIRAL SHOCKS WITH THERMAL INSTABILITY}
\shortauthors{KIM, KIM, \& OSTRIKER}
\begin{document}

\title{Galactic Spiral Shocks with Thermal Instability}

\author{Chang-Goo Kim\altaffilmark{1}, Woong-Tae Kim\altaffilmark{1},
and Eve C.\ Ostriker\altaffilmark{2}}
\affil{$^1$Department of Physics \& Astronomy, FPRD,
Seoul National University, Seoul 151-742, Republic of Korea}
\affil{$^2$Department of Astronomy, University of Maryland, 
College Park, MD 20742, USA}
\email{kimcg@astro.snu.ac.kr, wkim@astro.snu.ac.kr, ostriker@astro.umd.edu}
\slugcomment{Accepted for Publication in \apj}

\begin{abstract}
Using one-dimensional hydrodynamic simulations including 
interstellar heating, cooling, and thermal conduction,
we investigate nonlinear evolution of gas flow across galactic spiral
arms.  We model 
the gas as a non-self-gravitating, unmagnetized fluid, and follow
its interaction with a stellar spiral potential in a local frame 
comoving with the stellar pattern.
Initially uniform gas with density $n_0$ in the range 
$0.5\pcc\leq n_0 \leq 10\pcc$
rapidly separates into warm 
and cold phases as a result of thermal instability (TI), and also
forms a quasi-steady shock that prompts phase transitions.
After saturation, the flow follows a recurring cycle: 
warm and cold phases in the interarm region are 
shocked and immediately cool to become a denser cold medium in the
arm; post-shock expansion reduces the mean density
to the unstable regime in the transition zone and TI subsequently 
mediates evolution back into warm and cold interarm phases.
For our standard model with $n_0=2\pcc$, the gas resides in 
the dense arm, thermally-unstable transition zone, and interarm
region for 14\%, 22\%, 64\% of the arm-to-arm crossing time. 
These regions occupy 1\%, 16\%, and 83\% of the arm-to-arm distance,
respectively.  Gas at intermediate temperatures (i.e.\ neither warm
stable nor cold states) represents $\sim25$-30\% 
of the total mass, similar to the fractions estimated from \ion{H}{1}
observations (larger interarm distances could reduce this
mass fraction, whereas other physical processes associated with 
star formation could increase it).
Despite  transient features and multiphase structure, 
the time-averaged shock profiles 
can be matched to that of a diffusive isothermal medium
with temperature $1,000\Kel$ (which is $\ll T_{\rm warm}$) and 
``particle'' mean free path of $l_0=100\pc$.
Finally, we quantify numerical conductivity associated with translational 
motion of phase-separated gas  on the grid, and show that 
convergence of numerical results requires the numerical conductivity 
to be comparable to or smaller than the physical conductivity.
\end{abstract}

\section{Introduction}

Spiral arms are the most prominent features in disk galaxies.
As the interstellar medium (ISM) passes through the moderate
gravitational potential 
well of the stellar spiral arms, it is strongly compressed
and shocked, producing narrow dust lanes in optical images.
Active star formation is subsequently triggered in high-density
clouds inside the arms, resulting in downstream optical arms that 
contain OB associations and giant \ion{H}{2} regions 
distributed in a ``beads on a string'' fashion 
(e.g., \citealt{baa63,elm83,2006ApJ...642..158E,she07}). 
Other arm substructures include filamentary gaseous spurs (or feathers) 
seen in optical extinction, IR emission from dust, and
H$\alpha$ emission from star formation
(e.g. 
\citealt{1980ApJ...242..528E,scorec01,sco01,ken04,2004ApJS..154..222W,
lav06,gor07}), 
and giant molecular associations and atomic superclouds seen 
in CO and \ion{H}{1} radio observations
(e.g., \citealt{elm83,vog88,ran90,kna93}).
The locations of these arm substructures downstream from the primary dust
lanes indicates that the shock compression represents the first 
step in an evolutionary sequence that begins with diffuse ISM gas and
ends with star formation (for strongly-bound cores) 
and dispersal (for more weakly self-gravitating structures),
although it is uncertain whether spiral arms actually enhance 
star formation rate or just organize it
(e.g., \citealt{ger78,elm86,sle96,sei02}).

Studies of galactic spiral shocks date back to \citet{rob69}, 
who used a semi-analytic approach to 
obtain one-dimensional, stationary shock profiles 
as functions of the distance perpendicular to the shocks
(see also \citealt{fuj68,rob70,shu73}).  
\citet{woo75} used time-dependent calculations to show that spiral
shocks in local models indeed develop within one or two crossings 
of 
the background arm potential.  This and subsequent work
(e.g. \citealt{kim02}) suggests spiral arm 
shocks in the one-dimensional approximation are highly stable
for a range of the arm strength.  On the other hand, spiral shocks
have been shown to be intrinsically unstable when the vertical dimension is 
included \citep{mar98,gom02,gom04,bol06,kim06}. 
Since the arm-to-arm crossing periods are in general incommensurable 
with the vertical oscillation periods, the gas streamlines are not closed,
giving rise to shock flapping motions that dump a significant amount of 
random kinetic energy in the gas  \citep{kko06}.  Under certain
(strong compression) conditions, two-dimensional {\it in-plane} 
spiral shocks can also become unstable due to strong shear within the arm
\citep{wad04,dob06}.  However, these in-plane modes are stabilized by
moderate magnetic fields \citep{she06,dob08}, and suppressed in fully
three-dimensional models due to vertical dynamics \citep{kim06}.

Inclusion of gaseous self-gravity tends to enhance the arm response and 
symmetrize the density profile \citep{lub86}, and causes the shock front
to move downstream relative to the minimum in the potential \citep{kim02}.  
High post-shock density enhances the growth of self-gravitating
perturbations 
within spiral arms, although postshock flow expansion can limit this growth 
\citep{bal85,bal88}.
Using two-dimensional simulations with both self-gravity and magnetic 
fields, 
\citet{kim02} demonstrated that magneto-Jeans instability (in which magnetic
tension forces counterbalance the stabilizing Coriolis forces) leads
to the formation of both arm spurs and GMAs/GMCs with realistic properties
(see also \citealt{lyn66,elm94}). Subsequent studies including
three-dimensional effects \citep{kim06} and global spiral structure 
\citep{she06} have confirmed these findings.

While recent work has improved our understanding of
galactic spiral shocks and their larger 
substructures, these studies have oversimplified
the ISM thermodynamics, usually adopting 
an isothermal equation of state.
This ignores  potential consequences of thermal instability (TI) 
(\citealt{fie65}; see also \citealt{mee96} for review), which
 changes an otherwise homogeneous ISM to clumpy, multi-phase gas. 
(e.g., \citealt{fie69,hei01,wol03}).  In the classical
two-phase picture of the ISM,  cold dense clouds are in pressure equilibrium
with warm intercloud media that surround them \citep{fie69}.
Supernovae lead to a hot, diffuse third phase
\citep{1974ApJ...189L.105C,mo77}, but because massive star formation
is spatially
correlated and much of the hot gas produced is vented away, most of
the volume remains relatively unaffected (e.g. 
\citealt{1998ApJ...503..700F,2004A&A...425..899D}).
Since the cold clouds and the warm intercloud gas 
differ in density and temperature
by about two orders of magnitudes, their respective responses to 
spiral shocks and downstream expansion flows will be much different 
from the isothermal case.
When realistic thermal processes are considered, for instance,
warm rarefied gas in the interarm region can be converted 
via shocks  to cold dense gas in the arm regions. 
The reverse phase transition can then occur downstream for some
fraction of the mass, yielding  a quasi-steady cyclic exchange.

\citet{shu72} were the first to study the effects of gas cooling and 
heating on spiral shocks.  By considering a mixture of the comoving two 
stable phases and allowing for phase transitions, they calculated 
steady-state shock profiles for both cold and warm phases.  However, 
they employed a pressure-density relation, instead of solving 
the time-dependent energy equation, based on the assumption of
instantaneous thermal equilibrium; this precluded the possibility 
of unstable-phase gas in their calculations.

More recent years have seen 
a few numerical studies of spiral shocks with explicit heating and cooling, but
most of these suffer from strong numerical diffusion.
\citet{bak74} argued that allowance 
for thermal phase changes produces ``accretion fronts/waves'' 
instead of spiral shocks, in which the inflowing material radiates
its energy away.  As they mentioned, however, 
this result could  be due to large numerical diffusion;
we will indeed show below that in a moving medium, 
numerical conductivity can be large enough to 
suppress TI. On the other hand,
\citet{tub80} and \citet{mar83} have shown that some models
develop spiral shocks in which phase transitions from warm interarm gas 
to cold arm clouds occur, although insufficient resolution in their models
made the cloud sizes and separations significantly overestimated 
and prevented the transition regions from cold to warm phases 
from being resolved.
Very recently, \citet{dob07} and \citet{dob08} studied 
the effects of the warm phase on a pre-existing cold phase using
particle simulations, but they did not allow for phase transitions 
that are crucial in spiral shocks with TI.

In this paper, we initiate a study of galactic spiral shocks 
subject to ISM heating and cooling, using 
very high-resolution numerical hydrodynamic simulations.
We consider one-dimensional models that represent slices
perpendicular to the arm.  
We ignore the gaseous self-gravity and magnetic fields here,
deferring studies of these effects to future work.  
Our primary objectives are to determine overall shock structures under TI,
to explore where and how the transitions among the cold, warm, and 
unstable phases occur, and to find statistical properties 
such as temperature distributions, mass fractions, and velocity dispersions. 

The remainder of this paper is organized as follows: 
in \S2 we describe the basic equations we solve and present 
our model parameters and numerical methods.
In \S3 we test our numerical code and quantify the
diffusion (due to translational motion over the grid)
in terms of a numerical conductivity.
In \S4 we address the evolution  of cold and warm 
phases as they traverse spiral shocks, and provide statistical
measures to quantify the exchange cycle that develops. 
Finally, we summarize our results and discuss their implications in \S5. 

\section{Numerical Methods}

\subsection{Basic Equations}

We study  galactic gas flows and thermodynamic evolution in response
to an external
stellar spiral potential, which is assumed to be tightly wound
with a pitch angle $i\ll1$, and rotating at a constant pattern speed $\Omega_p$
with respect to an inertial frame.
For local simulations, it is advantageous to
set up a frame corotating with the spiral pattern, 
centered at the position $(R, \phi) =(R_0, \Omega_p t)$.
The local frame is tilted by an angle $i$ relative to the radial direction 
in such a way that the two orthogonal axes correspond to the directions 
perpendicular ($x$-axis) and parallel ($y$-axis) to the local arm segment, 
respectively \citep[][]{rob69}.
We assume that all physical variables depend only on the $x$-coordinate,
while allowing nonzero velocity in the $y$-direction.
Since the independent variable in our local models is a projection of 
the position on a streamline in a large-scale flow onto the $x$-axis,
the distance on the $x$-axis divided by $\sin i$ represents the 
distance that the flow has traversed in the azimuthal direction
along the streamline.  Therefore, the {\it temporal} interval between 
one arm crossing and the next crossing is the same as it would be for a 
global model. 

In the local arm frame, the background velocity due to galactic rotation 
is approximately given by
\begin{equation}\label{eq:v0}
 \vel_0=R_0(\Omega_0-\Omega_p)\sin i\xhat
+[R_0(\Omega_0-\Omega_p)-q_0\Omega_0 x]\yhat,
\end{equation}
where $\Omega_0$ is the angular velocity of gas at $R_0$ in the inertial frame
and $q_0\equiv-(d\ln\Omega/d\ln R)|_{R_0}$ is the local shear rate 
in the background flow in the absence of the spiral potential 
\citep[][]{kim02,kim06}.
Assuming that the motions induced by the stellar potential are much smaller 
than $R_0\Omega_0$, the basic equations of ideal hydrodynamics expanded in
the local frame read
\begin{equation}\label{eq:cont}
 \frac{\partial\rho}{\partial t}+\nabla\cdot(\rho \vel_T)=0,
\end{equation}
\begin{equation}\label{eq:mom}
\frac{\partial\vel_T}{\partial t}+\vel_T\cdot\nabla\vel_T=
-\frac{1}{\rho}\nabla P -q_0\Omega_0 v_{0x}\yhat-2\mathbf{\Omega}_0\times
\vel-\nabla\Phi_{\rm ext},
\end{equation}
\begin{equation}\label{eq:energy}
\frac{\partial e}{\partial t} +\vel_T\cdot\nabla e =
-\frac{\gamma}{\gamma-1}P\nabla\cdot\vel_T -\rho\mathcal{L}+\nabla\cdot(\mathcal{K}\nabla T),
\end{equation}
\citep[see][]{rob69, shu73, bal88, kim02,pio04}, 
where $\vel_T\equiv \vel_0+\vel$ is the total velocity in the local frame,
$\Phi_{\rm ext}$ is the external stellar spiral potential, 
$\rho\mathcal{L}(\rho, T)$ is the net cooling function,
and $\mathcal{K}$ is the thermal conductivity. 
Other symbols have their usual meanings. 
We adopt an ideal gas law $P=(\gamma-1)e$ with $\gamma=5/3$.

For the stellar spiral potential, we consider a 
simple sinusoidal shape:
\begin{equation}\label{eq:ephi}
 \Phi_{\rm ext}=\Phi_{\rm sp}\cos\left(\frac{2\pi x}{L_x}\right),
\end{equation}
analogous to a
logarithmic potential of \citet{rob69} and \citet{shu73}.
In equation (\ref{eq:ephi}), $\Phi_{\rm sp}$ denotes the amplitude of the 
spiral potential, while $L_x =2\pi R_0\sin i/m$ 
is the arm-to-arm separation for an $m$-armed spiral.  
We take the size of the simulation domain equal to $L_x$;
since $x$ varies from $-L_x/2$ to $L_x/2$ and $\Phi_{\rm sp}<0$, 
$\Phi_{\rm ext}$ attains its minimum at the center ($x=0$).
We parametrize the spiral arm strength using 
\begin{equation}\label{eq:F}
F\equiv \frac{m}{\sin i}\left(\frac{|\Phi_{\rm sp}|}{R_0^2\Omega_0^2}\right),
\end{equation}
which measures the maximum force due to the spiral 
potential relative to the the mean axisymmetric gravitational force 
\citep{rob69}.  

The net cooling function per unit volume is given by 
$\rho\mathcal{L}\equiv n(n\Lambda[T]-\Gamma)$, 
where $n=\rho/(\mu m_{\rm H})$ is the gas number density 
and $\mu=1.27$ is the the mean molecular weight per particle.
For the heating and cooling rates of the atomic ISM, 
we take the fitting formulae
\begin{equation}\label{eq:heat}
 \Gamma = 2.0\times10^{-26}\ergs,
\end{equation}
\begin{equation}\label{eq:cool}
 \frac{\Lambda(T)}{\Gamma}=10^7\exp\left(\frac{-1.184\times10^5}{T+1000}\right)
 +1.4\times10^{-2}\sqrt{T}\exp\left(\frac{-92}{T}\right) \cm^3,
\end{equation}
suggested by \citet{koy02} (see also \citealt{vaz07}).  
Under the adopted net cooling curve, the minimum and maximum
pressures for the coexistence of the classical warm/cold phases 
in a static equilibrium are 
$P_{\rm min}/k_B=1600\Kcm$ and $P_{\rm max}/k_B=5000\Kcm$.
The corresponding transition temperatures
$T_{\rm max}=5012\Kel$ and $T_{\rm min}=185\Kel$  
define cold ($T<T_{\rm min}$), warm ($T>T_{\rm max}$),
and intermediate-temperature phases ($T_{\rm min}<T<T_{\rm max}$).
To resolve the length scales of TI numerically 
(e.g., \citealt{koy04,pio04}), 
we include a constant value of thermal conductivity 
$\mathcal{K}_0=10^5\ergs\cm^{-1}{\, \rm K}^{-1}$.\footnote{While thermal 
conductivity is proportional to $T^{1/2}$ for neutral hydrogen at 
kinetic temperatures below $4.5\times 10^4\Kel$  \citep{par53},
we choose for simplicity a fixed value corresponding to 
thermal equilibrium at $T=1500\Kel$ for our standard density $n_0=2\pcc$.} 
The associated Field length below which thermal conduction 
erases temperature perturbations completely is defined by
\begin{equation}\label{eq:field}
 \lambda_F=2\pi\left\{\frac{\rho^2\Lambda}{\mathcal{K}_0T}
 \left[1-\left(\frac{\partial \ln \Lambda}{\partial \ln T}
 \right)\right]\right\}^{-1/2}
\end{equation}
\citep{fie65}.  In our models, $\lambda_F$ typically 
amounts to $\sim 0.18\pc$. 

\subsection{Model Parameters \& Numerical Methods}

We consider a simulation box in which the gas is initially homogeneous 
with density $n_0$ and pressure
$P_0$ when a spiral perturbation is absent.
Other than thermal processes involving TI, overall dynamics and
structures of spiral shocks in our models are completely characterized by 
the arm-to-arm distance $L_x$ and the flow speed $v_{0x}$ relative to 
the perturbing stellar potential in 
the $x$-direction (as well as $\sin i$, $q_0$, $F$, and $\Omega_p/\Omega_0$).
The azimuthal wavenumber $m$ of spiral arms is arbitrary,  and
the box location and the gaseous angular speed 
relative to the arms can then be specified as
$R_0=mL_x/(2\pi\sin i)$ and 
$\Omega_0-\Omega_p=2\pi v_{0x}/(mL_x)$, respectively.
To achieve the numerical resolution sufficient to resolve the Field length, 
we consider a small box with $L_x=628$ pc. 
For the relative velocity, we choose $v_{0x}=13\kms$  corresponding to 
the rotational velocity of $R_0\Omega_0=260\kms$ with a flat rotation curve 
($q_0=1$).  
The corresponding arm-to-arm crossing time is $\tcro\equiv L_x/v_{0x} = 
4.7\times10^7\yr$, which we choose as the fiducial time unit in our 
presentation.  
For spiral arm parameters, we take pattern speed $\Omega_p=\Omega_0/2$,
pitch angle $\sin i=0.1$, and strength $F=5\%$ in all the models. 
We note that driven by numerical requirement, 
spiral arms in our models have a small separation and thus a short 
dynamical time, so that some of our numerical results 
(e.g., mass fractions) that depend 
on the ratio of cooling time to dynamical time may not be 
applicable to spiral arms with a much larger arm-to-arm crossing time.

To simulate spiral shocks with varying total gas contents, 
we consider 12 models that have the same initial thermal pressure 
$P_0/k_B$ but differ in the initial density $n_0$;  
we adopt the solar neighborhood value at $P_0/k_B=3000\Kcm$ 
based on the observational (e.g., \citealt{fer01,hei03,jen07}) and theoretical 
(e.g., \citealt{wol03}) arguments.
Note that although some models are out of thermal equilibrium initially,
they immediately tend towards equilibrium owing to rapid
heating and cooling, with the equilibrium pressure
depending on $n_0$. 
Table~\ref{tbl:model} lists the model parameters and simulation outcomes.
Column (1) labels each run; while the models with the prefix SU
rapidly undergo TI even with $F=0$, the
SW and SC models would stay warm or cold throughout
were it not for the spiral perturbations. 
Column (2) lists $n_0$.  Columns (3)--(8) give the width
of, and time spent in, the arm, transition, and interarm regions,
respectively,  in each model. 
The mass and volume fractions of the cold, warm, and intermediate-temperature 
phases are given in columns (9)-(14), respectively.
We take model SU2 with $n_0=2\pcc$ as our fiducial model; at this density,
the total surface density would be $\sim 12\Msun\pc^{-2}$
for vertical scale height of $\sim100\pc$.

We integrate the time-dependent partial differential equations 
(\ref{eq:cont})--(\ref{eq:energy})
using a modified version of the Athena code \citep{gar05}. 
Athena implements a single step, directionally unsplit Godunov scheme for 
compressible hydrodynamics in multi-spatial dimensions and
allows a variety of spatial reconstruction methods and approximate Riemann 
solvers;
the version we use here employs the piecewise linear method (PLM) with 
the Roe Riemann solver.\footnote{
Although the piecewise parabolic method (PPM) is known to provide 
in general less diffusive spatial reconstruction than the PLM, 
our experiments have shown that the PPM often produced negative 
internal energy in the presence of strong radiative cooling
inside spiral shocks.} 
We implement the shearing-periodic boundary condition at the 
$x$-boundaries \citep{haw95}.
Because of the very short cooling time, energy updates from net 
cooling are made implicitly based on Newton-Raphson iteration,
while the conduction term is solved fully explicitly. 
For stable and accurate results, we ensure the time step is kept 
smaller than the CFL condition for thermal conduction as well as the 
cooling/heating time (see \citealt{pio04}).
Our standard models employ $N=16,384$ zones, corresponding to 
grid spacing of $\Delta x = 0.04\pc$, which satisfies the condition
$\Delta x < \lambda_F/3$ for convergence of numerical results 
\citep{koy04}; we also ran models with different
grid sizes in order to study the effects of numerical resolution. 

\section{Code Test and Numerical Conductivity\label{sec:numc}}

The Athena code we use has been verified on a wide variety of test 
problems including hydrodynamic shock tubes, advection of a square box
in a background shear flow, and one-dimensional propagation of sound waves
in a rotating, shearing medium.
For simulations involving cooling/heating and conduction terms, 
it is crucial to check if
the numerical scheme employed can resolve the length and times scales of 
the fastest growing TI modes.  This test is of particular importance
for the current work since the gas in our models is non-static, moving 
in the $x$-direction at an average speed of $v_{0x}=13\kms$, so that
 numerical diffusion associated with advection and zone averaging 
may reduce the growth rates of TI at small scales.
In this section, we describe the test results 
of our numerical code on the development of TI 
in static, rotating, and moving media 
in the absence of spiral potential perturbations, and 
provide a quantitative measure of the
numerical diffusion in terms of numerical conductivity.

\subsection{Linear Dispersion Relation}

We test our code by comparing the numerical growth rate of a 
particular TI mode with the corresponding analytic prediction.  
To this end, we derive a dispersion relation for local, axisymmetric TI 
in a homogeneous medium that is rotating, shearing, and undergoing uniform 
translational motion with $\vel_{0}$.  
We linearize equations (\ref{eq:cont})--(\ref{eq:energy}) 
(dropping the external potential), assuming 
plane-wave disturbances $\propto  e^{nt - i k x}$, 
where $n$ and $k$ are the growth rate and
wavenumber of the disturbances in the $x$-direction, respectively.
We follow the same steps as in \citet{fie65}, except that we include
the non-zero background motions and the indirect forces arising from 
galaxy rotation.  The resulting dispersion relation for local disturbances
is given by
\begin{eqnarray}\label{eq:disp}
  \tilde n^3   
+ \tilde n^2 a\left(k_T+\frac{k^2}{k_\mathcal{K}}\right) 
+ \tilde n(\kappa^2+k^2 a^2) 
+ a\left[\frac{a^2k^2}{\gamma}\left(k_T-k_\rho+\frac{k^2}{k_\mathcal{K}}\right)
   +\kappa^2\left(k_T+\frac{k^2}{k_\mathcal{K}}\right)\right]=0,
\end{eqnarray} 
where 
$\tilde n \equiv n - i kv_{0x}$ is the Doppler-shifted growth rate,
$a\equiv(\gamma k_B T/\mu m_H)^{1/2}$ is the adiabatic speed of sound,
$\kappa\equiv( R^{-3}d(R^4\Omega^2)/dR|_{R_0})^{1/2}=(4-2q_0)^{1/2}\Omega_0$
is the local epicyclic frequency,
and $k_T, k_\rho, k_\mathcal{K}$ are the wavenumbers defined by
\begin{eqnarray}
k_T=\frac{\gamma(\gamma-1)}{a^3} 
 \left(\frac{\partial \mathcal{L}}{\partial\ln T}\right)_\rho, \;\;\; 
k_\rho=\frac{\gamma(\gamma-1)}{a^3} 
 \left(\frac{\partial \mathcal{L}}{\partial\ln \rho}\right)_T, \;\;\;
k_\mathcal{K} = \frac{a^3\rho}{\gamma(\gamma-1)\mathcal{K} T},
\end{eqnarray}
(e.g., \citealt{fie65}).
In the limit of $q_0=0$ and $v_{0x}=0$, 
equation (\ref{eq:disp}) recovers the dispersion relation given by 
\citet{fie65} for TI in a rigidly-rotating medium with
no translational motion.  The fact that $v_{0x}$ occurs in the
dispersion relation only through $\tilde n$ implies that 
a constant translational motion does not change 
the growth rates of TI if measured in a frame moving with $v_{0x}$,
analogous to Galilean invariance in mechanics.

For isobaric TI to occur, the term in the square 
brackets in equation (\ref{eq:disp}) must be negative. 
This of course necessitates $k_T - k_\rho < 0$, the 
Field criterion for isobaric TI \citep{fie65}.
Thermal conduction and rotation suppress short and long wavelength 
perturbations against TI, respectively, reducing the unstable 
range of wavelengths to $k_1 < k < k_2$, where
$k_1$ and $k_2$ are two positive roots of 
$k^4 + (k_\mathcal{K}[k_T - k_\rho] + \gamma \kappa^2/a^2) k^2
+ \gamma \kappa^2k_Tk_\mathcal{K}/a^2=0$.
One can show that 
$k_2 \rightarrow (k_\mathcal{K}[k_\rho - k_T])^{1/2}=2\pi/\lambda_F$
for weakly- or non-rotating systems ($\kappa \rightarrow 0$), 
while $k_1^2 \rightarrow \gamma \kappa^2 k_T /(a^2[k_\rho - k_T])$
in the limit of vanishingly small conductivity 
($k_\mathcal{K} \rightarrow \infty$).
Figure~\ref{fig:grate} plots as various lines sample growth rates 
$\tilde n$ of TI calculated from equation (\ref{eq:disp})
for cases $\mathcal{K}=\Omega_0=0$ (\textit{dotted line}),
$\mathcal{K}=\mathcal{K}_0$ with $\Omega_0=0$ (\textit{solid line}),
and $\mathcal{K}=\mathcal{K}_0$  with $\Omega_0=130\kms\kpc^{-1}$ 
(\textit{dashed line}).
The density, pressure, and shear parameter are taken to be $n_0=2\pcc$, 
$P_0/k_B=3000\pcc$, and $q_0=1$, corresponding to our fiducial model SU2.
Stabilization of TI by rotation and conduction are apparent at large
and small scales, causing the wavelengths of the most unstable 
disturbances to occur at $\lambda\sim3\pc$ in between the
cut-off wavelengths.

\subsection{Numerical Conductivity} 

\begin{figure}
 \epsscale{1.}
 \plotone{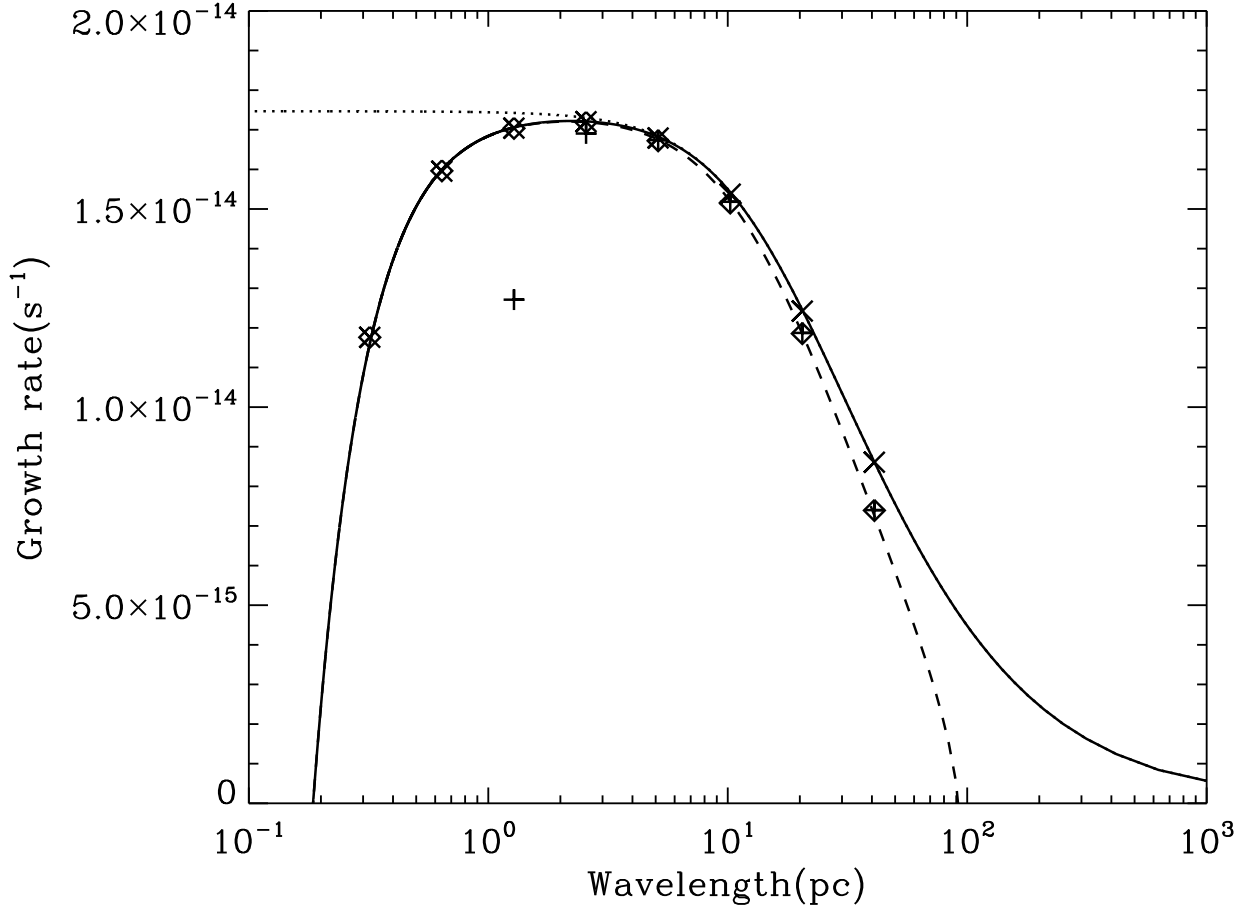}
 \caption{Analytic growth rate of TI versus perturbation wavelength
(eq.\ [\ref{eq:disp}]) for cases 
$\mathcal{K}=\Omega_0=0$ (\textit{dotted}), 
$\mathcal{K}=\mathcal{K}_0$ with $\Omega_0=0$ (\textit{solid}),
and $\mathcal{K}=\mathcal{K}_0$ with $\Omega_0=130\kms\kpc^{-1}$ 
(\textit{dashed}), showing that rotation and conduction stabilize
perturbations at large and small scales, respectively.  
Symbols indicate the numerical growth 
rates measured from test simulations, all with
$\mathcal{K}=\mathcal{K}_0$: for a static medium (\textit{cross}),
for a rotating medium without translational motion (\textit{diamond}), 
and for a medium with both rotational and translational motion
(\textit{plus}).  Note that numerical diffusion associated with 
translational motion over the grid reduces the growth rates at small scales
significantly.  See text for details.}
 \label{fig:grate}
\end{figure}

For our code tests, we consider three kinds of models depending
on $\Omega_0$ and $\Omega_p$:
(1) a static disk ($\Omega_0=\Omega_p=0$);
(2) a rotating disk without translational motion 
($\Omega_0=\Omega_p=130\kms\kpc^{-1}$, yielding $v_{0x}=0$
from eq.\ [\ref{eq:v0}]) 
(3) a rotating disk with translational motion 
($\Omega_0=2\Omega_p=130\kms\kpc^{-1}$, so that $v_{0x}=13\kms$).
The same conductivity $\mathcal{K}=\mathcal{K}_0$ is adopted
for all the models.
Other parameters including $n_0$, $P_0$, and $q_0$ are 
the same as in model SU2.
In each model, we initialize an eigenmode of TI with wavelength $\lambda$ 
and vary the box size to fit it in, while keeping the grid
spacing $\Delta x=0.04\pc$ fixed.  We monitor the evolution of the maximum
density and measure its growth rate numerically in the linear regime.
Figure~\ref{fig:grate} plots the resulting growth rates from 
the runs as cross, diamond, and plus symbols for the first,
second, and third types of models, respectively.
Evidently, the numerical and analytic results are in good agreement for 
models without the translational motion, confirming the performance of our
implementation of the heating/cooling and conduction terms.
While the translational motion of the gas at a level of $v_{0x}=13\kms$ 
does not affect large-scale modes much,
it significantly reduces the numerical growth rates for small-scale modes 
with $\lambda < 2\pc$.
The discrepancy between the numerical and analytic results 
appear to be due to numerical diffusion that causes the thermal energy 
to spread out as cooling/heating regions are advected 
with the background flow.

To quantify the strength of numerical heat diffusion in our code, 
we solve the linear dispersion relation (\ref{eq:disp}) 
for a given set of parameters to find the effective conductivity 
$\mathcal{K}_{\rm eff}$ that yields an analytic growth rate equal 
to the numerical value obtained from the model simulation with 
the same parameters except for the 
physical conductivity $\mathcal{K}=\mathcal{K}_0$.
We then calculate the numerical conductivity from
$\mathcal{K}_n=\mathcal{K}_{\rm eff}-\mathcal{K}$.
In the case of models with $v_{0x}=13\kms$ 
and $\lambda=2.56$ and 1.28 pc shown 
in Figure~\ref{fig:grate}, for instance,
$\mathcal{K}_n=3.4\times10^{5}$ and
$1.2\times10^{6}\ergs\cm^{-1}\Kel^{-1}$, 
respectively,
about 3.4 and 12 times larger than $\mathcal{K}_0$.

\begin{figure}
 \epsscale{1.}
 \plotone{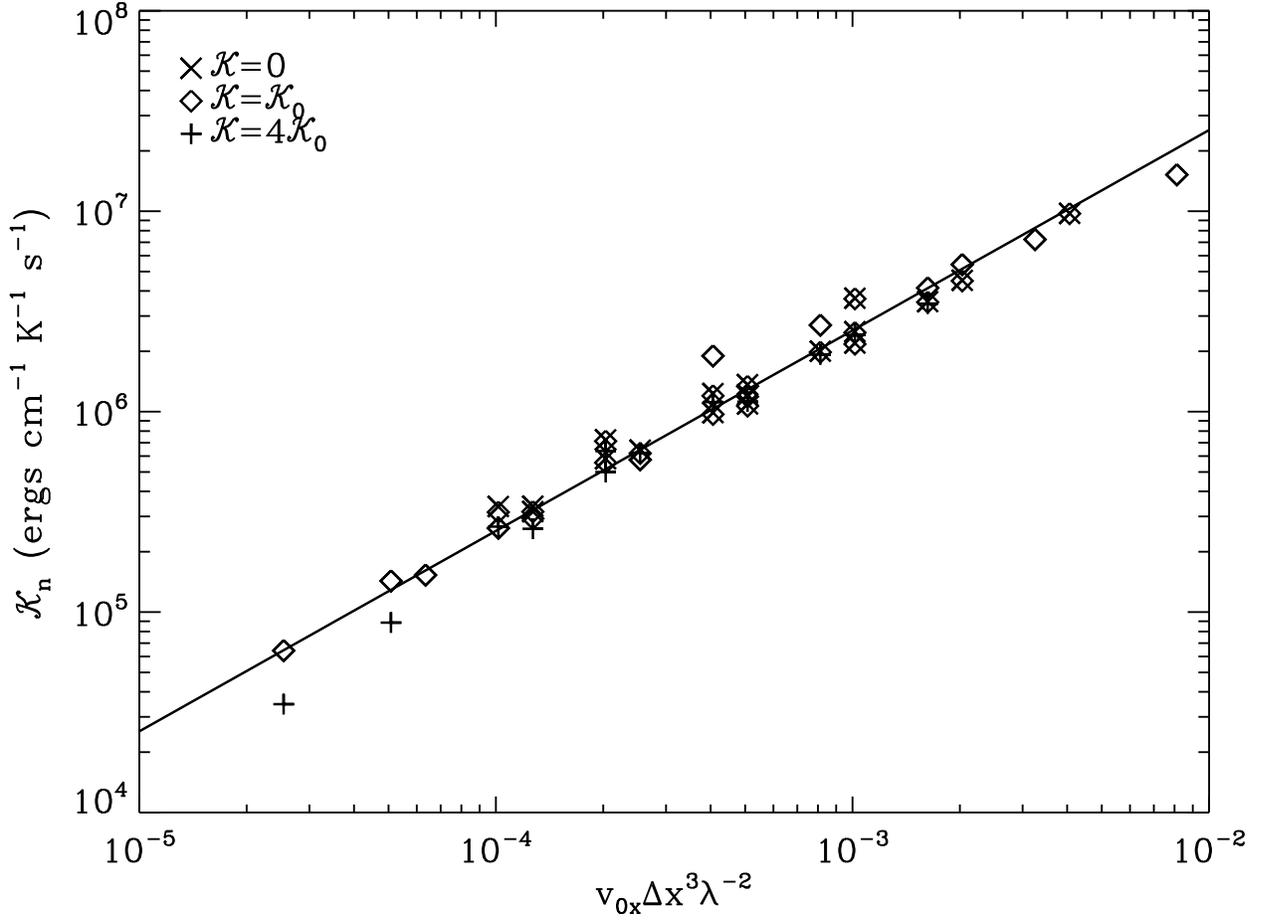}
 \caption{Dependence of numerical conductivity $\mathcal{K}_n$ on
the translational velocity $v_{0x}$, the grid size $\Delta x$,
and the perturbation wavelength $\lambda$ of TI, from
the test simulations with physical conductivity 
$\mathcal{K}=0$ (\textit{cross}), 
$\mathcal{K}=\mathcal{K}_0$ (\textit{diamond}),
and $\mathcal{K}=4\mathcal{K}_0$ (\textit{plus}).
Note that $v_{0x}$ is expressed in units of
$\kms$, while $\Delta x$ and $\lambda$ are in parsecs.
The solid line, represented by equation (\ref{eq:numk}), 
gives the best fit to the test results based on 
the second-order PLM for spatial reconstruction in the Athena code
we use.
}
 \label{fig:knum}
\end{figure}

In order to find the parametric dependences of $\mathcal{K}_n$ on 
the translational velocity, grid size, and perturbation wavelength,
we ran a suite of models varying $v_{0x}$ from 
2.6 to $26\kms$, $\Delta x$ from 0.02 to $0.16\pc$,
and $\lambda$ from 0.32 to $10.2\pc$.  
We also explored the cases with different levels of physical conductivity at
$\mathcal{K}=0$, $\mathcal{K}_0$, or $4\mathcal{K}_0$.
The numerical conductivity is calculated only when 
the numerical growth rate is lower than the analytic value at
the same $\mathcal{K}$ by more than 1\%.
Figure~\ref{fig:knum} plots as various symbols the resulting numerical
conductivity based on the PLM scheme for spatial reconstruction,
showing that $\mathcal{K}_n$ is well fitted by
\begin{equation}\label{eq:numk}
 \mathcal{K}_n=10^{9.4}\left(\frac{v_{0x}}{1\kms}\right)
                       \left(\frac{\Delta x}{1\pc}\right)^3
                       \left(\frac{\lambda}{1\pc}\right)^{-2}
\ergs\cm^{-1}{\, \rm K}^{-1}.
\end{equation}
This suggests that the numerical conductivity in a moving medium 
depends rather sensitively on the numerical resolution. 
Equation (\ref{eq:numk}) states that our models with $\Delta x=0.04\pc$
have numerical conductivity comparable to the physical conductivity
for the fastest growing modes,
so that the effect of numerical diffusion on the results presented
in this paper is not significant.

We repeated the same calculations using the higher-order PPM
reconstruction scheme, and found
the numerical conductivity is linearly proportional to 
$v_{0x}\Delta x^{4}\lambda^{-3}$.
Generalizing these results,  we rewrite equation (\ref{eq:numk}) as
$\mathcal{K}_n\propto v_{0x}\Delta x(\Delta x/ \lambda)^{p}$, 
where $p$ is the order of the spatial reconstruction scheme 
($p=2$ and $3$ for PLM and PPM, respectively).  This implies that 
the numerical conductivity can be viewed as 
the diffusion coefficient ($\propto v_{0x}\Delta x$) modified by the
accuracy of the interpolation scheme used in spatial reconstruction 
($\propto[\Delta x/\lambda]^{p}$).

\section{Nonlinear Simulations}

\subsection{Standard Model\label{sec:evol}}

We now study the nonlinear evolution of thermally-unstable gas flows 
under an imposed spiral potential.  In this subsection, we 
focus on model SU2 with $n_0=2\pcc$.  We initially apply density
perturbations created by a Gaussian random field with a
power spectrum $|\rho_k|^2 \propto k^{-5/3}$ for 
$1 \leq 2\pi k/L_x \leq 128$ 
and zero power for $2\pi k/L_x>128$ in the Fourier space, 
corresponding to a one-dimensional Kolmogorov spectrum for sonic disturbances. 
The standard deviation of the density perturbations is fixed to 
be 1\% in physical space.
In order to suppress rapid motions of the gas 
caused by an abrupt introduction of the spiral potential,
we slowly turn it on, reaching the full level $F=5\%$ at $t/\tcro\sim2.4$.

\begin{figure}
 \epsscale{1.0}
 \plotone{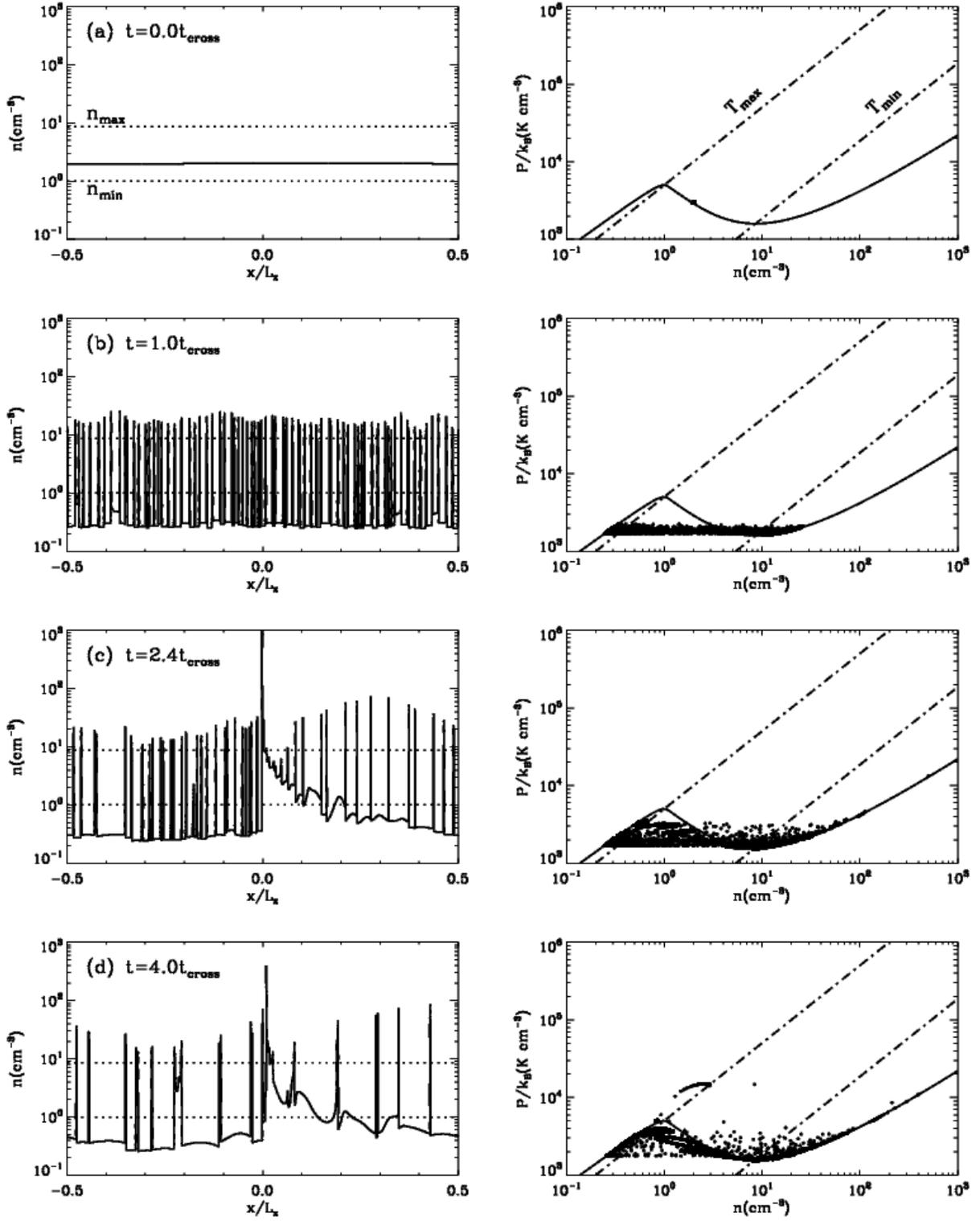}
 \caption{\textit{Left}: 
Density snapshots at 
$t/\tcro=0.0$, $1.0$, $2.4$, $4.0$ from model SU2.
Dotted lines labeled by $n_{\rm max}$ and $n_{\rm min}$ demarcate
the cold and warm phases.
\textit{Right}: Corresponding scatter plots of $n$ versus $P/k_B$ 
together with the equilibrium cooling curve (\textit{solid line}).
Dot-dashed lines marked by $T_{\rm max}$ and $T_{\rm min}$ 
divide the warm ($T>T_{\rm max}$), intermediate-temperature
($T_{\rm min}<T<T_{\rm max}$), and cold ($T<T_{\rm min})$ phases.
 }
 \label{fig:tevol}
\end{figure}

Figure~\ref{fig:tevol} shows density distributions of model SU2 
at $t/\tcro=0$, 1, 2.4, and 4.0 together with scatter plots of pressure 
versus density overlaid on the equilibrium cooling curve.
The gas initially has $T=1500\Kel$ and is thus thermally unstable
(Fig.\ \ref{fig:tevol}a).  It evolves rapidly to 
cold ($n>n_{\rm max}=8.6\pcc$ and $T<T_{\rm min}=185\Kel$) and 
warm ($n<n_{\rm min}=1\pcc$ and $T>T_{\rm max}=5012\Kel$)
phases within $0.5\tcro$.  The TI soon saturates, 
and random cloud motions aided by epicyclic shaking
cause cold clouds to collide and merge together, or split sometimes, 
resulting in, on average, 70 cold clouds with a mean cloud separation 
of $\sim 0.8\pc$, which are   
in rough pressure equilibrium with the surrounding
warm gas at $P/k_B=1900\Kcm$ (Fig.\ \ref{fig:tevol}b).  
At this time, the spiral potential remains weak and  
most gas in the unstable temperature range corresponds to the boundaries 
of the cold clouds.  
As the amplitude of the potential grows, the gas is gathered 
toward the potential minimum, forming a spiral shock near $x=0$. 
The shock reaches maximum strength at around $t/\tcro=2.5$ 
shortly after $F$ attains the full strength.
Both cold and warm phases in the interarm regions continually enter the shock 
front and are compressed to become cold gas with higher density.  
They subsequently 
expand and become thermally unstable as they leave the spiral arm regions, 
returning back to the cold and warm interarm phases (see below). 
At about $t/\tcro=3$, the overall shock structure reaches a quasi-steady 
state in the sense that the mass and volume fractions of each phase do
not change appreciably with time, although the shock oscillates slightly
around an equilibrium position and cold clouds shift as they follow 
galaxy rotation (projected onto the $\hat x$ direction). 

\begin{figure}
 \epsscale{1.0}
 \plotone{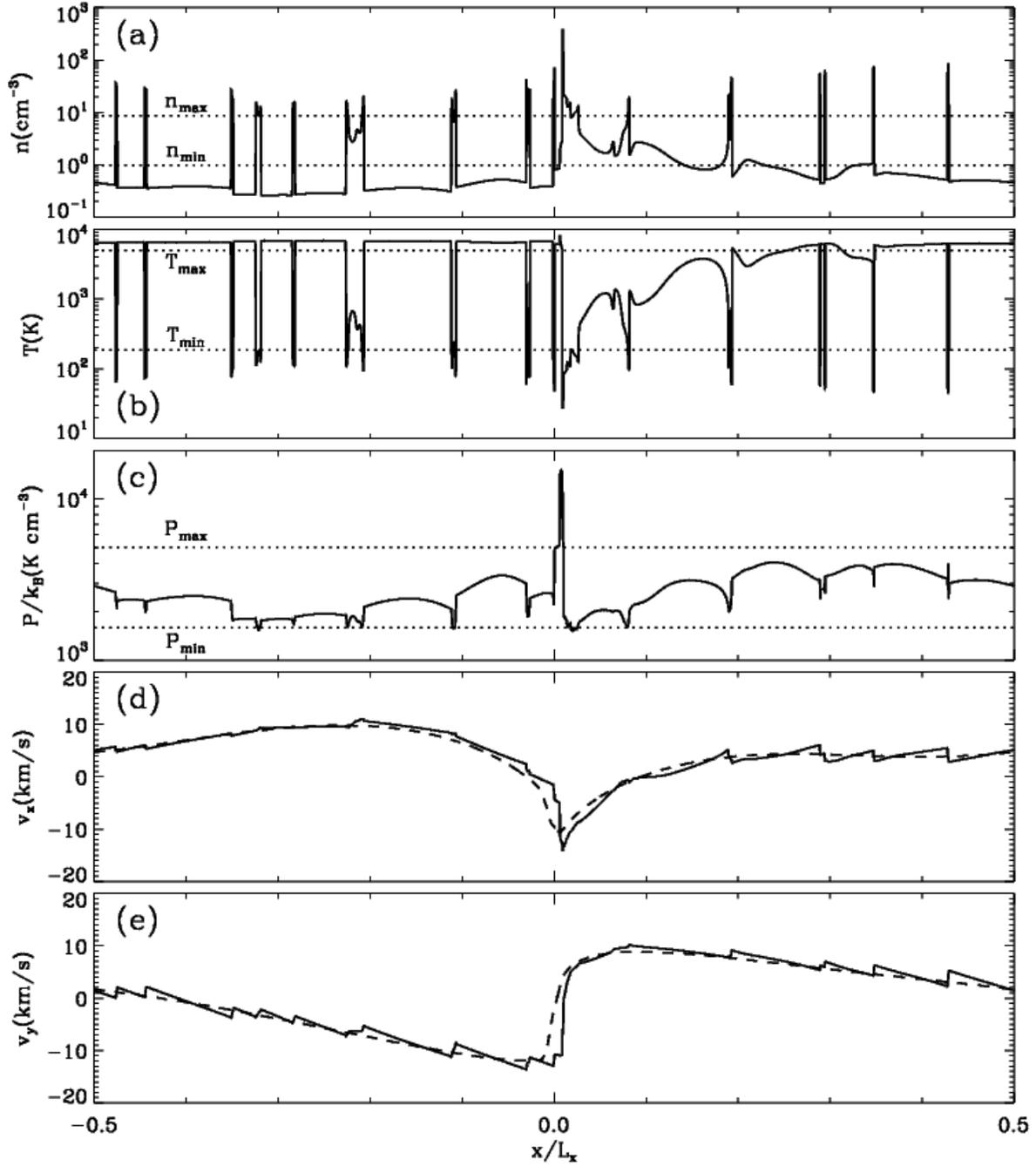}
 \caption{Profiles of number density, temperature, 
pressure, and perturbation components of the 
velocities perpendicular ($v_x = v_{T,x}-v_{0x}$) 
and parallel ($v_y=v_{T,y}-v_{0y}$) to the arm,
for  model SU2 at $t/\tcro=4.0$.
In (a) to (c), dotted lines demark the transitions between
the cold, intermediate-temperature, and warm phases.
Dashed lines in (d) and (e) show the time-averaged 
velocity (streaming) profiles over $t/\tcro=5-8$.
 }
 \label{fig:sat}
\end{figure}

\begin{figure}
 \epsscale{1.}
 \plotone{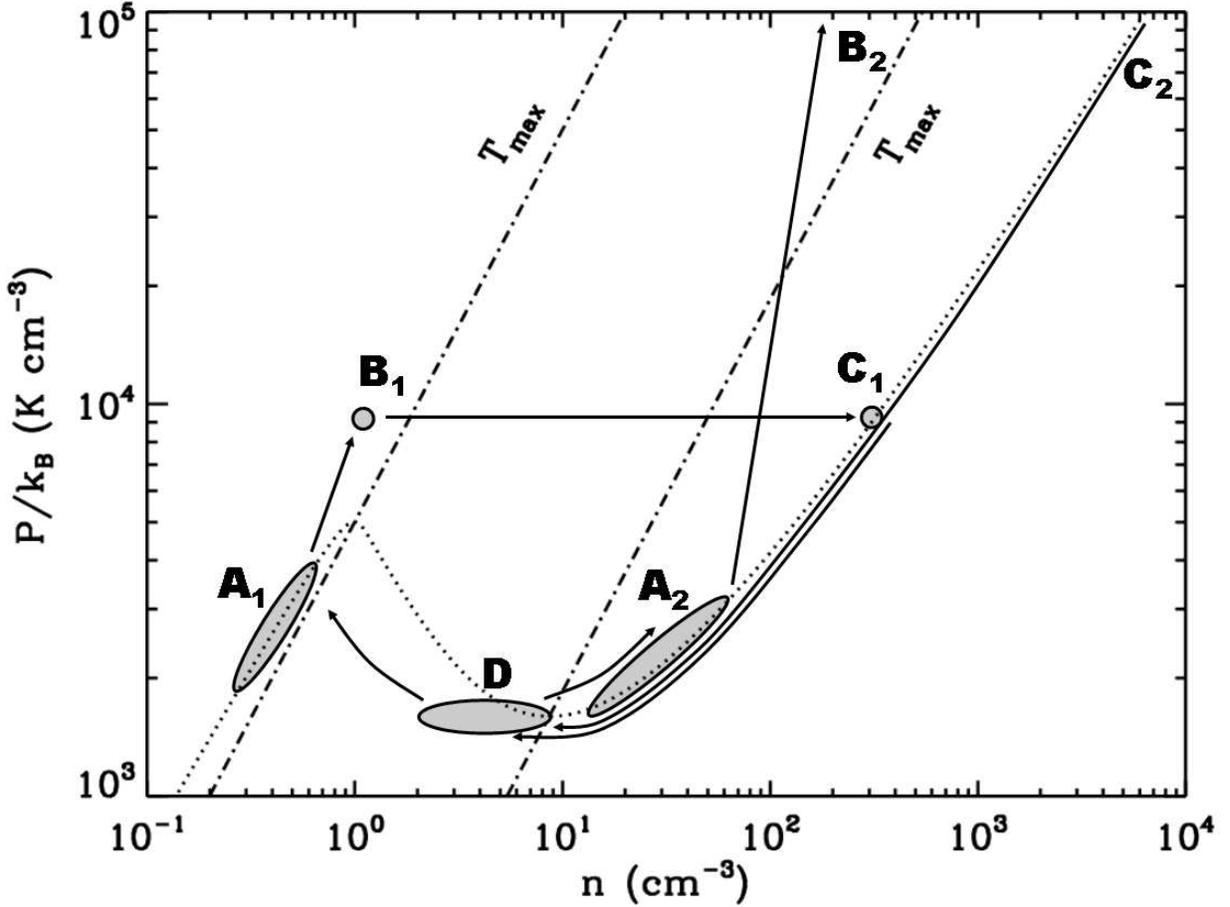}
\caption{Schematic evolutionary track in the $n-P$ plane (\textit{solid lines
with arrows}) of gas cycling through a spiral pattern with TI.
The dotted line is the thermal equilibrium curve, 
while dot-dashed lines mark $T_{\rm max}$ and $T_{\rm min}$.
The warm and cold phases in the interarm regions (A) are shocked
(B), undergo radiative cooling (C), become thermally unstable 
due to postshock expansion (D), and return to the 
interarm two-phase state again (A).  See text for details.}
 \label{fig:schematic}
\end{figure}

Figure~\ref{fig:sat} plots the distributions
of physical variables in model SU2 at $t/\tcro=4$ after this quasi-steady
state has been reached; many spikes and discontinuities in density as well as 
sawtooth-like velocity profiles are evident.
Figure~\ref{fig:schematic} schematically illustrates the evolutionary
tracks of the cold and warm phases in the $n$--$P$ plane.
Regions marked with A, B, C, and D correspond to interarm, 
immediate post-shock, spiral arm, and thermally-unstable 
transition zones, respectively;
the transition zone refers to the window downstream
from the arm (between $x/L_x\sim0.03-0.2$) in Figure \ref{fig:sat}b, 
where most of the gas has temperatures in the unstable range between
$T_{\rm min}$ and $T_{\rm max}$.
The subscripts 1 and 2 denote the warm and cold phases, respectively,
of interarm gas, and their respective immediate post-shock counterparts.

In the interarm regions (either $x/L_x<0$ or $x/L_x>0.2$
in Fig.\ \ref{fig:sat}), 
warm (A$_1$) and cold (A$_2$) phases are in rough equilibrium 
in terms of the total (= thermal + ram) pressure.
While the cold interarm clouds typically have slightly lower thermal
pressure than the warm interarm medium, their ram pressure is of
comparable magnitude, i.e. $P/k_B\sim 2000-4000\Kcm$. 
Some of the cold interarm 
clouds exhibit ``double-horned'' 
structure, a consequence of merging with neighbors or splitting into
two pieces (see e.g., Fig.\ \ref{fig:sat}a).
As the warm and cold interarm phases enter the shock front, they experience 
a strong compression, jumping to B$_1$ and B$_2$, respectively.
While the density jump is by a factor of $4$ for both cold and warm phases, 
the pressure jump for the cold phase is about 10 times larger than
in the warm phase since the former is about $100$ times colder.\footnote{
Strictly speaking, this holds true only for one-dimensional shocks.
In two or three dimensions, small clouds can experience enhanced 
compression as the shock wrapping around them is able to propagate 
towards the cloud centers. They may also be subject to dynamical effects 
such as Kelvin-Helmholtz and Rayleigh-Taylor instabilities 
(e.g., \citealt{woo76}).}
The shocked gas in states B$_1$ and B$_2$ is not in thermal balance
and subsequently undergoes strong post-shock cooling
(e.g., \citealt{muf74}), moving almost 
isobarically to C$_1$ and C$_2$. The transition from A to C is
essentially instantaneous (cooling time $\sim 10^3$ yrs).\footnote{
Because of its very large post-shock density, the transition 
from A$_2$ to C$_2$ occurs over an extremely short cooling scale
($\sim 0.02$ pc) that is not resolved in our models.}

One of the characteristics of galactic spiral shocks is that gas
accelerates after the maximum shock compression 
(e.g., \citealt{rob69,bal88,kim02}).
Because of this post-shock expansion, the cold gas inside the arm 
($0<x/L_x<0.03$) 
becomes progressively less dense, evolving from C to D.  
Since the dynamical time scale is much longer than the cooling time 
(due to the reduced velocity inside the arm), the cold gas either
at C$_1$ or C$_2$ moves all the way down to the transition zone D 
 following  the equilibrium curve in the $n$--$P$ plane.  
When the expanding gas reaches region D ($0.03<x/L_x<0.2$), it
becomes thermally unstable and turns back into either the warm (A$_1$) 
or cold (A$_2$) interarm phase.
For model SU2, it typically takes about
$\sim 0.14\tcro$ from C to D stages and $\sim 0.22\tcro$ from D to A stages.
Gas is in the interarm regime A for the balance of the cycle
($\sim0.64\tcro$).
When averaged over $t/\tcro=5-8$, 
the arm, transition, and interarm regions in model SU2 occupy 
approximately 1, 16, 83\% of the spatial domain, respectively. 

\begin{figure}
 \epsscale{1.0}
 \plotone{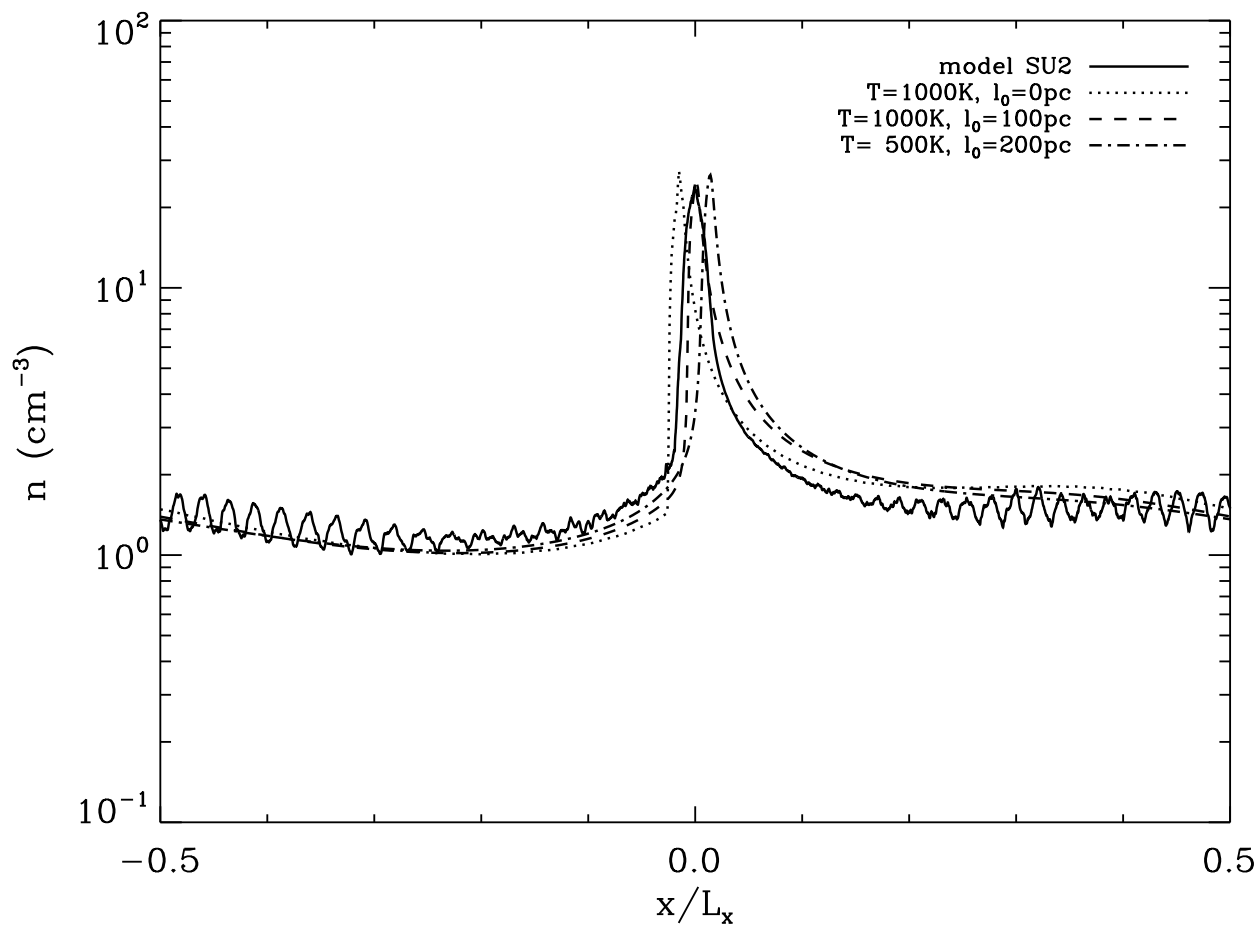}
 \caption{Density profile averaged over
$t/\tcro=5-8$ from model SU2 (\textit{solid}), compared to
stationary shock profiles for isothermal models. \textit{Dotted curve}:  
$T=1000\Kel$ non-viscous model; 
\textit{dot-dashed curve}: $T=500\Kel$ 
viscous models with $l_0=200\pc$; 
\textit{dashed curve}:
$T=1000\Kel$ viscous model with $l_0=100\pc$.
 }
 \label{fig:vis}
\end{figure}

Our models, as well as the real ISM, contain many interarm
clouds which would be too small to be detected individually in 
extra-galactic radio observations.
In modeling gaseous  spiral shocks,
\citet{lub86} did not explicitly solve for the cloudy structure.
Instead, they adopted an isothermal equation of state and treated 
the effects of cold clouds (and their
collisions) by including a viscous term parameterized by a mean
free path $l_0$ for the fluid.
They found that viscosity renders arm profiles smoother 
and more symmetric. 
In order to compare our models 
with a  single-phase viscous counterpart, 
we take a temporal average over $t/\tcro=5-8$ of the density 
distributions in model SU2. We then take a boxcar average 
of the time-averaged profile, with a window of $8$ pc.
Figure~\ref{fig:vis} plots the resulting mean density profile 
as a solid line.  For comparison, we have run isothermal models that 
include an explicit viscosity term in the momentum equation (\ref{eq:mom}) 
in a manner similar to in \citet{lub86}.  Selected results are shown 
in Figure~\ref{fig:vis}.
Evidently, larger viscosity tends to produce 
a weaker spiral shock, with a peak that is more symmetric and 
farther downstream.
In terms of the strength and placement of the spiral arm, 
an isothermal model with $T=1000\Kel$ and $l_0=100\pc$ provides a
fairly good match for the mean density profile when TI is included. 
The isothermal viscous model, however, has a slightly broader arm
compared to the time average of the multiphase model.
Our results thus demonstrate that (for the purposes of obtaining a
lower-resolution ``beam averaged'' profile) the effects of clouds 
embedded in a warm medium can be effectively modeled by viscosity
\citep{cow80,gam96}, but only if the medium's 
temperature and the mean free path are chosen appropriately.
We note that if a temperature comparable to that of the warm medium
were adopted for an isothermal counterpart, the shock would generally
be too weak.

\subsection{Effects of Initial Number Density\label{sec:ndep}}

\begin{figure}
 \epsscale{1.}
 \plotone{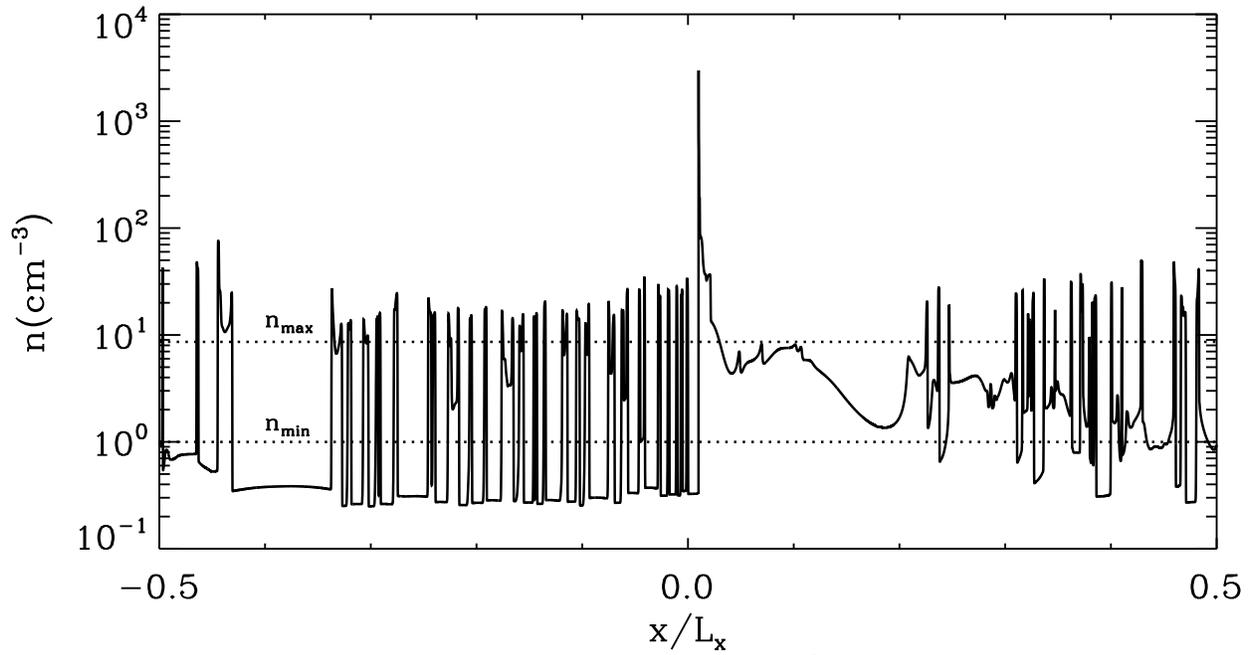}
 \caption{
Density profile of model SU5 with $n_0=5\pcc$ 
at $t/\tcro=4$.   Note that
the postshock transition zone in model SU5 is wider than that of
model SU2 shown in Figure~\ref{fig:sat}.
}
 \label{fig:high_n}
\end{figure}

All the SU models we consider start from a thermally-unstable initial state.
Since the TI growth time is short compared to the time over which we
turn on the spiral potential, these models 
first evolve into a thermally-bistable
state before developing shocks.  
Figure~\ref{fig:high_n} plots a density profile 
at $t/\tcro=4$ in model SU5 with $n_0=5\pcc$.
Comparing to model SU2 as shown in Figure~\ref{fig:tevol},
Figure~\ref{fig:high_n} shows that the densities (and temperatures) 
of the cold and warm phases after TI saturates
are insensitive to the initial gas density, 
although of course 
models with higher $n_0$ produce more  cold clumps.\footnote{ 
Mass conservation requires the number of cold clouds $N_c$ to be 
given approximately by $N_c\approx n_0 L_x/(ln_c)$, 
where $l$ and $n_c$ are the mean size and number density of the clouds. 
From simulations with differing $n_0$, we
empirically found that $l\sim 0.8\pc\;(n_0/2\cm^{-3})^{0.3}$,
i.e. the mean cloud size (presumably set by merging
and splitting), depends only 
weakly on $n_0$.  This gives $N_c \propto n_0^{0.7}$.}
The ensuing development of spiral shocks and evolutionary tracks in
the $n$--$P$ plane  are 
also qualitatively similar to those described in the previous subsection.

Perhaps the most notable difference in the late-time states
with different $n_0$ is the size of the postshock transition zone.
Columns (3)-(8) of Table~\ref{tbl:model} show that 
the transition zone (and also the arm region, to a lesser extent)
widens with increasing $n_0$.  One can easily see this by 
comparing Figures~\ref{fig:sat}d and \ref{fig:high_n}.
This trend is of course because models with larger $n_0$
have a larger fraction of the total mass in the cold phase that
evaporates to maintain an unstable state.
In addition, the mean separation of cold clouds inside the arm is smaller in 
models with higher $n_0$, allowing for
merging with neighboring clouds more often during the postshock
expansion stage. This reduces the expansion rate effectively, 
thereby extending the size of zones with unstable density
toward downstream.  

\begin{figure}
 \epsscale{1.}
 \plotone{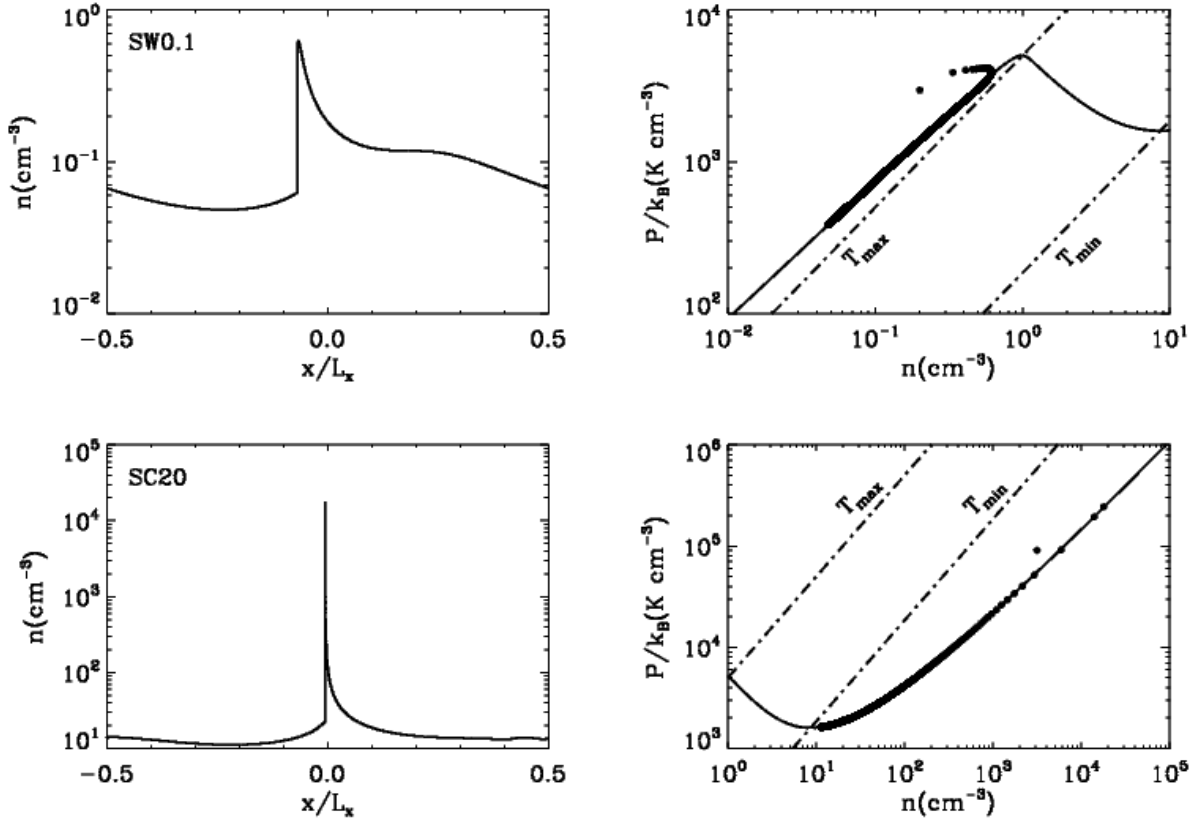}
 \caption{Density profiles and scatter plots in the
density--pressure plane of models SW0.1 with $n_0=0.1\pcc$ 
(\textit{upper frames}) and 
SC20 with $n_0=20\pcc$ (\textit{lower frames}) 
at $t/\tcro=2.4$.
Note that the spiral shocks in these models are smooth and 
each case contains only a single phase of gas.  The shock compression is
much stronger for all-cold than all-warm gas, because of the higher
Mach number.}
 \label{fig:single}
\end{figure}

Unlike the SU models, 
the SW and SC models initially have a low- or high-enough density
that they are thermally stable and the 
 early development of a spiral shock takes place in a 
single-phase medium.  Nevertheless, the shock compression and/or
post-shock expansion can still produce thermally unstable gas 
if the initial density is not far from the unstable range. 
For example, the shock in model SW0.5 drives 
the thermal pressure  above $P_{\rm max}$, and the gas evolves rapidly
via TI into  a cold stable phase
inside the arm. 
In model SC10, on the other hand, strong post-shock expansion 
drives the pressure of the initially cold gas below $P_{\rm min}$,
after which a 
fraction of the gas expands to become a warm phase. 
Consequently, the resulting structures at late times in models 
SW0.5 and SC10 are similar to those in the SU models.
In models with $n_0$ further away from the unstable values, however, 
changes in the gas pressure due to shock compression and  
postshock expansion are insufficient to trigger phase transitions. 
Figure~\ref{fig:single} plots equilibrium shock profiles in 
models SW0.1 and SC20, which start with $n_0=0.1$ and $20\pcc$, respectively.
The gas  in both of these models consists only of the warm or cold phase
and the profiles are smooth everywhere except at the shock front.

\subsection{Temperature Distribution}\label{sec:temp}

\begin{figure}
 \epsscale{1.}
 \plotone{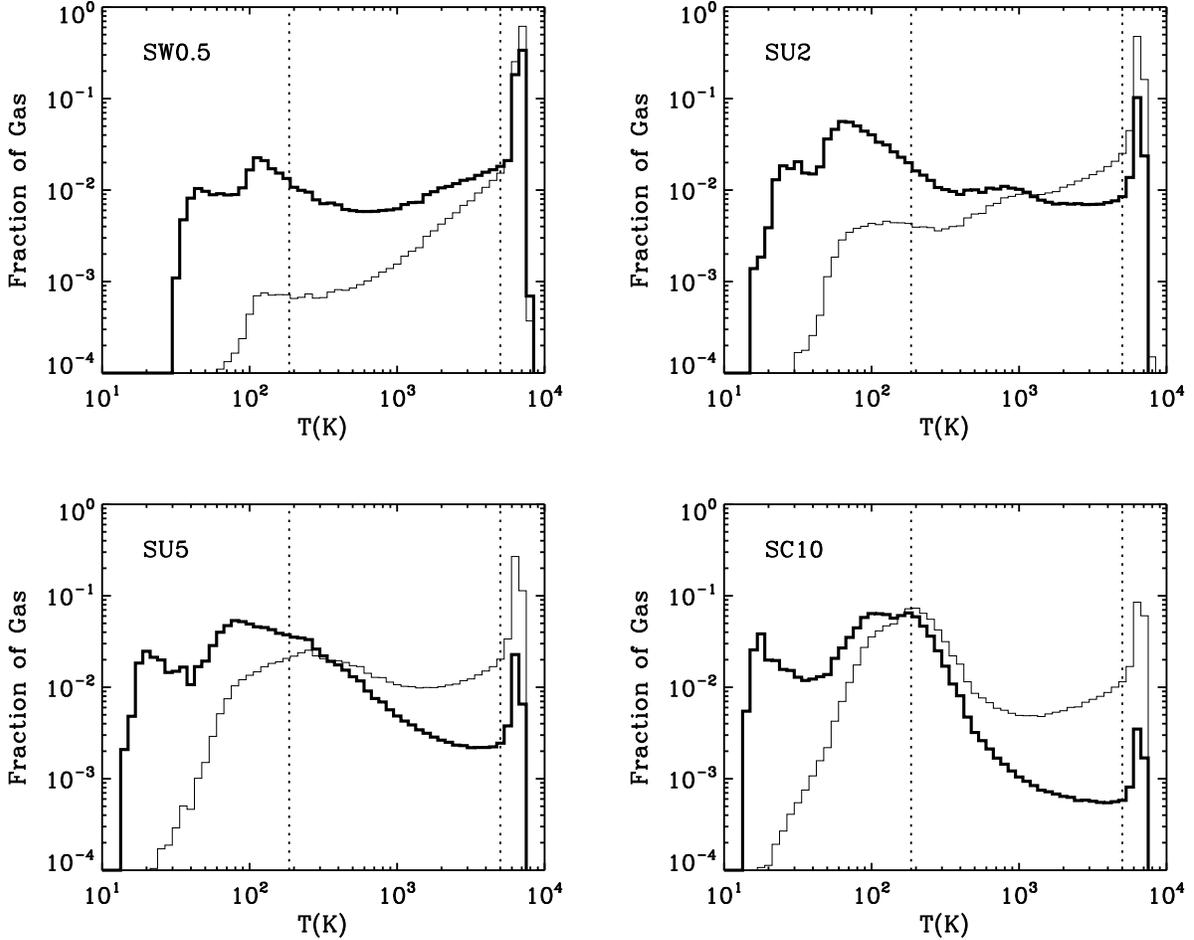}
 \caption{Mass-weighted ({\it thick lines}) 
and volume-weighted ({\it thin lines}) temperature PDFs 
averaged over $t/\tcro=5-8$ for models 
SW0.1, SU2, SU5, and SC10.  In each panel, vertical 
dotted lines indicate $T_{\rm min}$ and $T_{\rm max}$ to demark 
the cold, intermediate-temperature, and warm phases.
Of the cold component, the gas with $T\simlt 50\Kel$ is located in 
the arm region, while cold clouds with $50\Kel\simlt T\simlt T_{\rm min}$
are found in the interarm or transition zones.
 }
 \label{fig:pdf}
\end{figure}

Figure~\ref{fig:pdf} plots the volume-weighted and mass-weighted 
temperature probability distribution functions (PDFs) 
for models SW0.5, SU2, SU5, and SC10, averaged over $t/\tcro=5-8$. 
The vertical dotted lines in each panel indicate 
$T_{\rm min}$ and $T_{\rm max}$, 
marking the cold, intermediate-temperature, and warm phases.
For the standard model SU2,
the mass-weighed temperature PDF is characterized by 
a broad cold peak at $T\sim 20-150\Kel$ and a narrow warm peak 
at $T\sim 6000-8000\Kel$, although there also exists a substantial 
amount of the intermediate-temperature gas. 
In the broad cold peak, the portion of gas with $T\simlt 50\Kel$
corresponds to the very dense arm population immediately behind the shock,
while the portion with $50\Kel \simlt T < T_{\rm min}$ 
includes both arm and interarm cold clouds.
Since a larger initial density implies a larger initial cold fraction
even before
the spiral potential is applied, the cold mass
fractions in both arm and interarm regions increase in models with 
larger $n_0$. 
For model SW0.5, on the other hand,  
introduction of the spiral potential allows only a small fraction 
of the gas to exceed $P_{\rm min}$ and undergo TI,
resulting in a lower cold peak and a higher warm peak than in model SU2.
For all the models, the cold peak is fractionally much broader
(i.e. larger $\Delta T/T$) than the warm peak.

Of the total intermediate-temperature phase, about $70\%$ is found in the 
post-shock transition zone for models with TI,
while the remaining portion resides in boundary layers between the cold and 
warm phases in the interarm region.
Figure~\ref{fig:pdf} shows that the
distribution of the intermediate phase is almost flat for model
SU2 and increasingly favors lower temperature as $n_0$ increases.  
This is because models with larger $n_0$ have a slower post-shock expansion 
rate, and thus more gas near  $T_{\rm min}$.
The mean density and density-weighted mean temperature of the 
intermediate-temperature gas are found to scale as 
$n_i \approx 0.41 n_0 + 1.29$ and $T_i \approx 4458/(n_0 + 1.81)\Kel$
for $0.5 \simlt n_0 \simlt 10$, where $n_i$ and $n_0$ are in units 
of $\pcc$.  For the cold and warm phases, the mean values 
are $n_c\approx23\pcc, T_c\approx86\Kel$, and $n_w\approx0.39\pcc, 
T_w\approx6400\Kel$,
almost independent of $n_0$. These points lie slightly above $P_{\rm min}$
on the cold and warm branches of the thermal equilibrium curve.

\subsection{Mass and Volume Fractions\label{sec:mf}}

\begin{figure}
 \epsscale{1.}
 \plotone{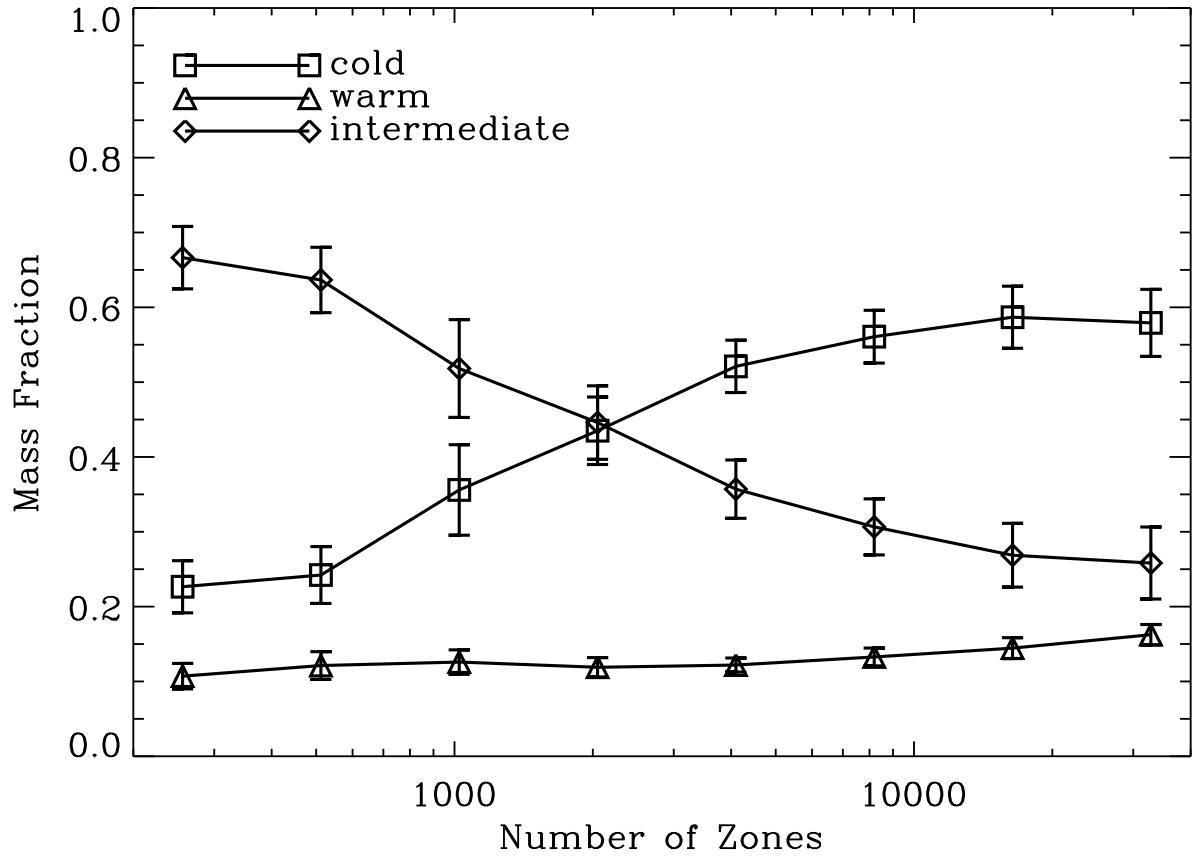}
 \caption{Mass fractions of the cold, intermediate-temperature,
and warm phases in model SU2 versus numerical resolution.
The numerical results are not affected by numerical conductivity
as long as the number of zones is larger than $10^4$.
 }
 \label{fig:conv}
\end{figure}

For models with  thermal instability,  we calculate the mass and volume 
fractions of three phases averaged over $t/\tcro=5-8$.  
Figure~\ref{fig:conv} plots against numerical resolution the mean 
mass fractions in model SU2, along with the standard deviations as errorbars. 
While the warm mass fraction is fairly insensitive to the number of
grid points $N$, the cold (intermediate-temperature) 
mass fraction increases (decreases) with increasing resolution for 
$N \simlt 10^4$.  This is mainly because of the numerical conductivity 
associated with a large zone size for small $N$, 
as explained in \S\ref{sec:numc}.  When numerical conductivity is large,
the boundary layers in between the cold and warm phases thicken in 
proportion to the Field length.  This increase in the
intermediate-temperature mass (in the interarm) occurs primarily 
at the expense of the cold phase \citep{beg90,fer93,ino06}.
In addition,  large numerical conductivity tends to suppress 
TI in the post-shock transition zone, yielding less cold gas. 

Overall, the broadening of interarm cloud interfaces in low resolution models
is responsible for 80\% of the increase in the intermediate-temperature mass,
while the remaining 20\% is due to the suppression of TI in the post-shock
transition zone.  There is no interarm cold gas in models with $N\simlt 512$,  
while arm cold gas still exists in these lowest resolution models since
the post-shock regions are, when fully resolved, denser and broader than 
interarm clouds, and thus are less affected by zone averaging.
As long as $N \simgt 10^4$, 
the numerical conductivity becomes comparable to, or
smaller than, the physical conductivity, and the results are
numerically well resolved.
Because most of the difference between moderate and high resolution
models is at warm/cold interfaces, the extremely high resolutions 
that we find are needed for accurate measurement of the
thermally-unstable mass fraction would not be required for studies
that focus primarily on shock dynamics.
We find that the overall shock structure (in terms of the breadth of
the transition zone, and the time-averaged profiles) are comparable
to those in the converged models provided $N\simgt 10^3$.  At these
moderate resolutions, edges of individual clouds are smeared out, but the
physically-important transition zone downstream from the arm is still
recovered. 

In the case of model SU2, the converged values of the mass fractions
are 59\%, 14\%, and 27\% for the cold, warm, and 
intermediate-temperature phases, respectively.  
By volume the cold takes up 4\% of the total, 
while the warm and intermediate-temperature media occupy 70\% and 26\%. 
Since these proportions are 80\%, 10\%, and 10\% by mass
and 10\%, 80\%, and 10\% by volume when the spiral potential is weak
or absent (e.g., Fig.\ \ref{fig:tevol}b), this
implies that spiral shocks and subsequent post-shock expansion 
zone are important for populating the 
intermediate-temperature portion of the phase plane.

\begin{figure}
 \epsscale{1.}
 \plotone{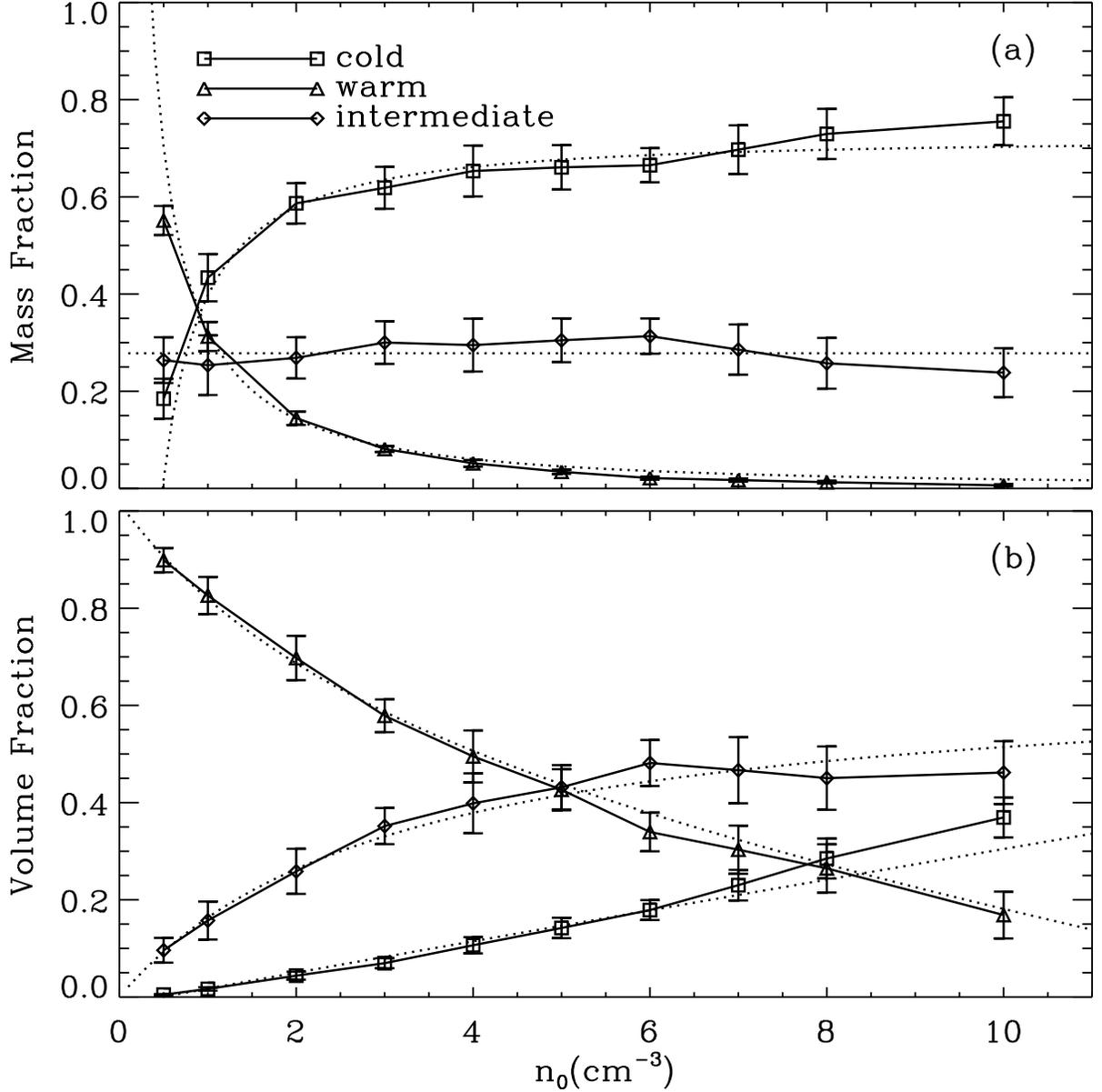}
 \caption{Mass and volume fractions of the cold (\textit{square}),
warm (\textit{triangle}), and intermediate-temperature (\textit{diamond}) 
phases averaged over $t/\tcro = 5-8$,
as functions of the initial number density $n_0$. 
Errorbars indicate the standard deviations of the measurements.
Dotted lines show the theoretical estimates computed by adopting a
constant value $f_i=0.28$ for the intermediate-temperature of the 
mass fraction.
 }
 \label{fig:mf_ndep}
\end{figure}

Columns (9)--(14) of Table~\ref{tbl:model} give the converged values 
of the mass and volume fractions of each phase in models 
with multiple phases.  
Figure~\ref{fig:mf_ndep} plots these proportions as functions of the initial 
number density $n_0$.  Interestingly,
the fraction of mass in the intermediate-temperature phase 
has a substantial value of 
$f_i \approx 0.28$, almost independent of $n_0$.\footnote{
By running models with differing $F$ (not listed in Table~\ref{tbl:model}),
we found that $f_i\approx 0.040 + 0.045 F(\%)$ 
for $3\%\leq F \leq 7\%$, but is insensitive to $n_0$ for the fixed 
arm parameters.}
The volume fraction of the intermediate phase increases with $n_0 \simlt 6\pcc$ 
and becomes flat at large $n_0$.  
Assuming that the cold, warm, and intermediate-temperature phases 
in each model are represented by the characteristic densities 
$n_c$, $n_w$, and $n_i$, respectively, mass conservation requires the mass 
fractions in the cold and warm phases to be 
$f_c = (1-n_w/n_0) - (1-n_w/n_i)f_i$ and
$f_w = n_w/n_0 - (n_w/n_i)f_i$, respectively,
where $n_c/n_w \gg 1$ is assumed (cf.\ \citealt{pio04}).
Dotted lines in Figure~\ref{fig:mf_ndep} plot these theoretical 
$f_c$ and $f_w$ using the empirical results $f_i=0.28$, $n_w=0.39\pcc$, and   
$n_i = 0.41 n_0 + 1.29\pcc$. These estimates are overall 
in good agreement with the simulation results.

\subsection{Velocity Dispersions\label{sec:veld}}

\begin{figure}
 \epsscale{1.}
\plotone{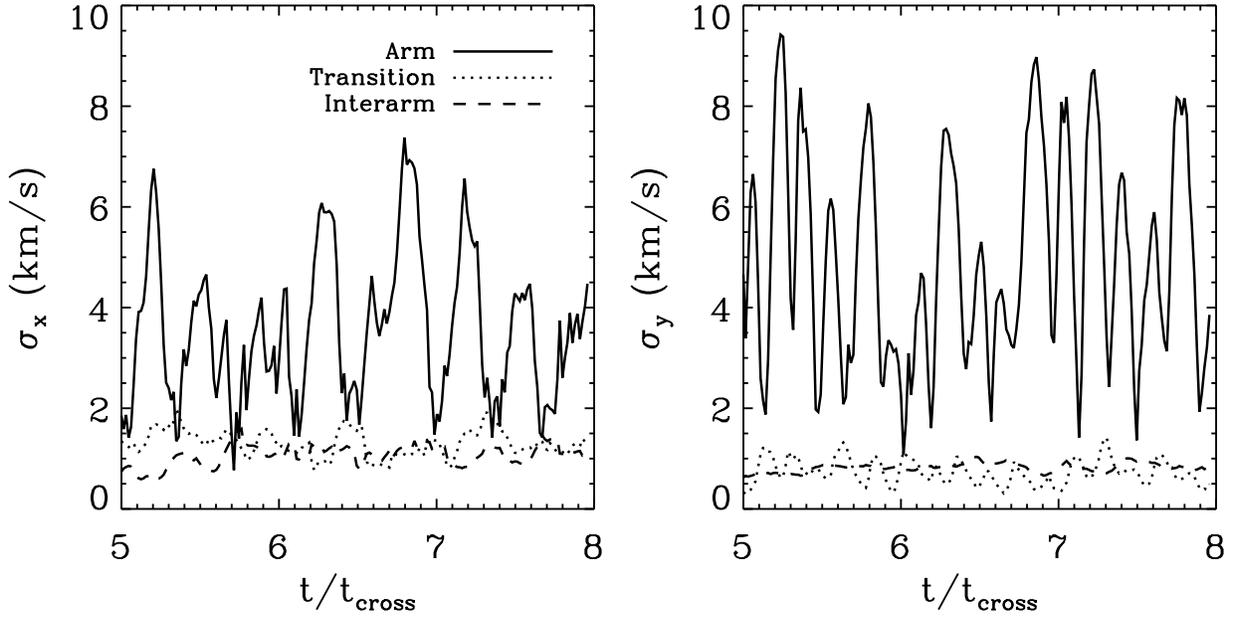}
\caption{Density-weighted velocity dispersions $\sigma_x$ and $\sigma_y$, 
relative to time-averaged template values, of the gas in the arm 
(\textit{solid}), transition (\textit{dotted}), interarm (\textit{dashed}) 
regions of model SU2.  The large-amplitude fluctuations of 
the velocity dispersions
in the arm region are caused
by incomplete subtraction of the arm streaming motions.  
The velocity dispersions in both unstable and interarm regions
are not subject to this effect, and have mean values of $\sim1.3\kms$
and $\sim 0.8\kms$ in the $x$- and $y$-directions, respectively.
 }
 \label{fig:veld_tevol}
\end{figure}

\begin{figure}
 \epsscale{1.}
\plotone{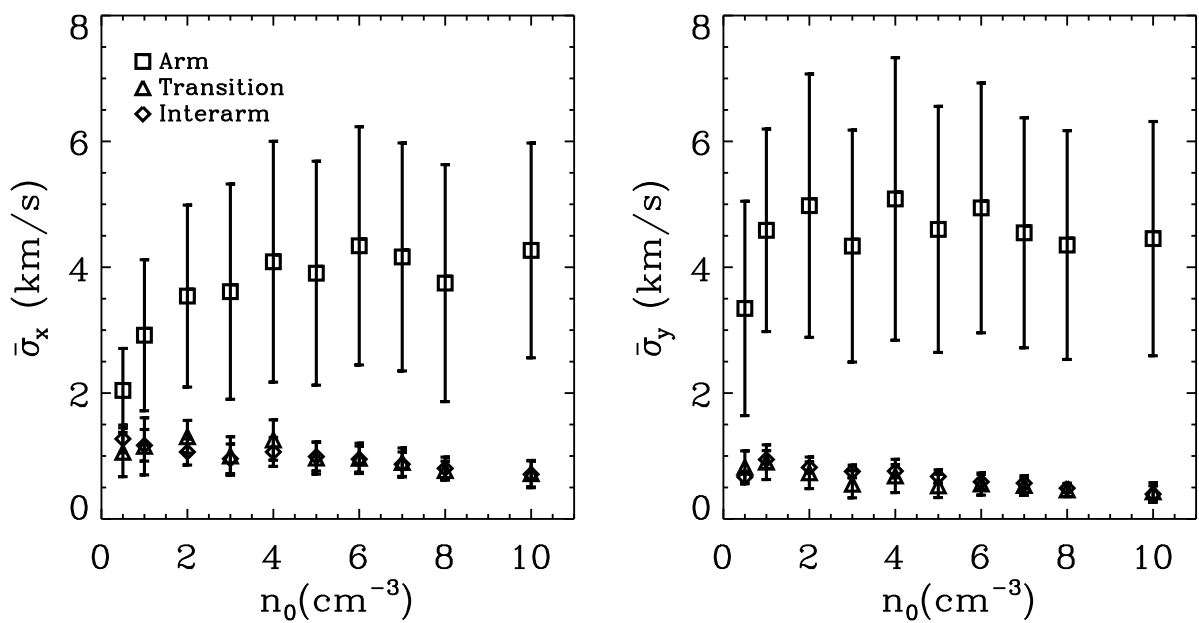}
 \caption{Mean values (\textit{symbols}) and standard 
deviations (\textit{errorbars}) of the density-weighted velocity 
dispersions of each phase during the time span $t/\tcro=5-8$ 
for models with multi-phase spiral shocks.
 }
 \label{fig:veld_ndep}
\end{figure}

Finally, we quantify the level of random gas motions driven in our models
due to TI and spiral potential perturbations.
Spiral arms produce gas streaming  motions that are ordered but vary
perpendicular to the shock front.  Since streaming velocities are much
larger in amplitude than the true random motions, care is needed in
measuring the latter.  In addition, while the overall shock profiles reach
a quasi-steady state after $t/\tcro=5$, several other effects make it
difficult to measure turbulent amplitudes in multiphase models:
randomly-placed cold clouds move with varying 
speeds in the interarm region, new structures are continuously developing 
in the transition zone, and the spiral shocks themselves
undergo small-amplitude oscillations perpendicular to the 
shock front. 
To separate out the background streaming from the total velocity 
as cleanly as possible, 
we first construct velocity template profiles $\langle v_x \rangle$ and 
$\langle v_y \rangle$ for each model, where 
the angle brackets denote a time average over
$t/\tcro=5-8$.
Dashed lines in Figure~\ref{fig:sat}d,e show sample
$\langle v_x \rangle$ and $\langle v_y \rangle$ profiles in model SU2.
We then calculate the density-weighted velocity dispersions using
$\sigma_i^2\equiv \int \rho(v_i-\langle v_i\rangle)^2dx /\int\rho dx$
(with $i=x$ or $y$) in the arm, interarm, and postshock transition zones
separately.
Figure~\ref{fig:veld_tevol} plots $\sigma_i(t)$ for model SU2,
while Figure~\ref{fig:veld_ndep} draws
the mean values $\bar\sigma_i \equiv \langle \sigma_i^2 \rangle^{1/2}$ 
along with the standard deviations $\Delta \sigma_i$ 
for models with multi-phase spiral shocks.

Figure~\ref{fig:veld_tevol} shows that the velocity dispersions of
the cold phase in the arm region exhibit large-amplitude temporal 
fluctuations, with characteristic periods of $\sim 0.4\tcro$ and $\sim0.2\tcro$
for $\sigma_x$ and $\sigma_y$, respectively.  
These variations of $\sigma_x$ and $\sigma_y$ are due to 
small-amplitude oscillations offsetting the spiral  shock front 
 from its  mean position. 
Since streaming varies strongly within the arm (Fig. \ref{fig:sat}d,e), 
the offset of the shock position leads to large differences between
instantaneous and mean streaming velocities within the arm; these
contaminate the measured velocity dispersion at locations near the
shock front.  
During the quasi-periodic oscillations of the spiral shock front,
$\sigma_x$ attains its minimum value when the shock is maximally displaced
upstream, while
$\sigma_y$ is smallest when the instantaneous shock 
front coincides with its mean position. 
Therefore, the local minima in the time series of 
$\sigma_x$ and $\sigma_y$ correspond to 
the upper limits to the level of random gas motions inside the arm.
For model SU2, the average values of the local minima in 
$\sigma_x(t)$ and $\sigma_y(t)$ are $\sim 2$ and $\sim 3\kms$, respectively,
which roughly equals  $\bar\sigma_i - \Delta\sigma_i$. As 
Figure~\ref{fig:veld_ndep} shows, $\bar\sigma_i - \Delta\sigma_i$ 
does not vary much with $n_0$, suggesting that the true velocity
dispersion within the arm is relatively independent of mean density. 

Unlike their behavior in the arm region, the time-averaged velocity 
profiles in the interarm and transition zones are relatively smooth, so that 
the velocity dispersions as calculated 
in these regions are relatively  free of
streaming motions. As Figures~\ref{fig:veld_tevol} and 
\ref{fig:veld_ndep} show, the velocity dispersions in the interarm and 
transition zones are similar, 
amounting to $\bar\sigma_x\sim1.3\kms,\bar\sigma_y\sim0.8\kms$ 
for model SU2, and decreasing slowly with increasing $n_0$.
The ratios of the velocity dispersions in the $x$- to $y$-directions
for all regions are consistent with predictions from epicyclic analysis
(e.g., \citealt{bin87}),
$\bar\sigma_y^2/\bar\sigma_x^2 = \kappa^2/(4\Omega_0^2) = 1 - q/2$.  Here,
the local shear rate
$q\equiv 2 - (2-q_0)n/n_0$ is modified from the background value 
$q_0\equiv -d\ln\Omega_0/d\ln R$, due to 
the constraint of potential vorticity conservation 
\citep{gam96,gam01,kim01,kim02}. 
For a flat rotation curve ($q_0=1$), 
local shear is reversed inside the arm where the local density exceeds
$2n_0$; this in turn increases 
$\bar\sigma_y /\bar\sigma_x \approx (n/2n_0)^{1/2}$ above $1/\sqrt{2}=0.7$. 
On the other hand, the whole interarm region and most of the unstable 
region have $n<n_0$, and thus the ratio 
$\bar\sigma_y /\bar\sigma_x$ is smaller than the prediction for a disk
without spiral structure.

\section{Summary \& Discussion}

\subsection{Summary}

Galactic spiral shocks in disk galaxies play an important role in 
structural and chemical evolution  by forming 
spiral-arm substructures and triggering star formation. 
Spiral shocks inherently involve large variations 
in the background density, 
while cooling and heating processes that determine the ISM density and
temperature 
depend rather sensitively on the local state.  The interplay between
these processes
may significantly alter
the shock strengths and structures, 
compared to those computed under an isothermal approximation. 
In particular,  large-scale compressions and expansions
across spiral arms may trigger
TI, thereby 
regulating transitions among the different ISM phases.
In this paper, we have used high-resolution
numerical simulations to
investigate the dynamics and thermodynamics of this highly nonlinear
process.  Our models include heating and cooling terms appropriate for
atomic gas 
explicitly in the energy equation, and thus naturally 
allow for transitions among cold, warm, and intermediate-temperature phases. 
The current investigation employs a one-dimensional model in which
all the physical quantities vary only in the in-plane 
direction perpendicular to
a local segment of a spiral arm. We allow for 
gas motions parallel to the arm, and include 
galactic differential 
rotation.  The effects of magnetic fields and self-gravity are neglected
in the present paper.  

Our main results are summarized as follows:

1. Background flow over the grid, represented by $v_{0x}$ in our
models, may result in a significant level of numerical diffusion  
in  finite-difference schemes dealing with the cooling/heating and conduction 
terms if the numerical resolution is not high enough.
We quantify the numerical diffusion of our implementation 
in terms of an effective numerical conductivity, and measure it using
growing modes of the thermal instability.  By comparing growth rates
with predictions from the linear dispersion relation including conduction,
we find that the numerical diffusion in the Athena code we use behaves 
as $\mathcal{K}_n\propto (v_{0x}\Delta x)(\Delta x/\lambda)^{p}$, 
where $\Delta x$ is the grid width, $p$ is the order of 
spatial reconstruction method ($p=2$ for PLM and 3 for PPM), and
$\lambda$ is the spatial wavelength.
For typical values of
$v_{0x}=13\kms$, $\Delta x=0.04\pc$, and $\lambda = 3.5\pc$
in our models, the numerical conductivity
with the PLM amounts to $\mathcal{K}_n 
\sim 2.3\times 10^{5}\ergs\cm^{-1}\Kel^{-1}$, 
comparable to the physical conductivity 
$\mathcal{K}_0 =10^5 \ergs\cm^{-1}\Kel^{-1}$
adopted in the current work.

2. Stellar spiral potential perturbations induce shocks that reach a
quasi-steady state after a few orbits.  The resulting flow contains
phase transitions provided the mean gas density is in the range 
$0.5\pcc \leq n_0 \leq 10 \pcc$, for the parameters considered 
in the present work.  Models with $n_0 \leq 0.1\pcc$ or $n_0 \geq 20\pcc$
yield single-phase spiral shock profiles.
We divide the flow into three distinct zones based on thermal regime: 
arm, interarm, and transition.
The ``arm'' refers to the highly-compressed postshock region filled with 
cold gas at $T<T_{\rm min}=185\Kel$, while in the ``interarm'' 
region far from the shock most of the volume is occupied 
by warm gas with $T>T_{\rm max}=5012\Kel$.  The ``transition'' zone
corresponds to the expanding region downstream from the arm where  
intermediate-temperature gas ($T_{\rm min} <T<T_{\rm max}$) undergoes TI.
Figure~\ref{fig:schematic} summarizes the evolutionary cycle in the 
density--pressure plane: the warm/cold interarm gas (A$_1$/A$_2$) is
shocked (B$_1$/B$_2$) and immediately cools 
to become the denser cold arm gas (C$_1$/C$_2$); this 
subsequently enters the unstable transition zone (D)  
and evolves back into the warm and cold interarm phases. 
For our standard model SU2 with mean density $n_0=2\pcc$,
the duration of the arm, transition, and interarm stages are 
approximately 14\%, 22\%, and 64\% of the cycle, 
respectively, 
occupying roughly 1\%, 16\%, and 83\% of the simulation domain.

3.  At late times, instantaneous profiles in models with TI 
show many  density spikes representing 
cold clouds. Time-averaged profiles
are, however, relatively smooth, and the density peaks representing
the arm are more symmetric than in non-diffusive, isothermal models.
We find that a viscous isothermal model with 
$T=1,000\Kel$ and mean free path of $l_0=100\pc$ yields a similar peak
density and arm width to the average profile from 
model SU2.  This confirms the notion that for purposes where detailed
ISM knowledge is not needed, multiphase effects 
on shocks can be approximately treated via a viscosity 
modeling excursions and collisions of dense clouds, as suggested by 
\citet{cow80} and \citet{lub86}.  Simulations such as those we have
performed are needed in order to calibrate the viscous/isothermal
model parameters, however.

4.  For models with multi-phase spiral shocks, 
 intermediate-temperature gas amounts to 
$\sim 0.25-0.3$ of the total by mass, insensitive to $n_0$.  
Of this, about 70\% is found in the transition zone, 
while the remaining 30\% lies at interfaces between the cold and
warm media.  This suggests that the postshock expanding flows, an
inherent feature of galactic spiral structure, is 
important for producing intermediate-temperature gas.  
The mean density and density-weighted mean temperature of the 
intermediate-temperature phase are found to 
be $n_i=0.41 n_0+1.29\pcc$ and $T_i = 4458/((n_0/1\pcc) + 1.81)\Kel$.
The fractions of the cold and warm phases 
are 59\% and 14\% by mass and 4\% and 70\% by volume for
model SU2, respectively, and vary with $n_0$ according to simple
expectations based on mass conservation with a prescribed density in
each phase.

5. We find that one-dimensional spiral shocks with multi-phase gas 
produces non-negligible random
gas motions.  At late times, the gas in both interarm and 
transition zones has 
typical 
density-weighted velocity dispersions of 
$\sigma_x\sim1.3\kms$ and $\sigma_y\sim0.8\kms$  in the directions
perpendicular and parallel to the spiral arms, respectively. 
This is trans-sonic with respect to the cold medium, and subsonic with
respect to the warm medium.
The cold gas in the arms is estimated to 
have slightly larger values $\sigma_x\sim 2\kms$ and 
$\sigma_y\sim 3\kms$, although true turbulence levels may be slightly
lower, because it is difficult to fully subtract streaming motions 
in the arm region.

\subsection{Discussion}

Since the growth rates of pure TI are maximized at the smallest available
wavelengths, growth of grid-scale noise dominates numerical
simulations if it is not suppressed.  To prevent this numerical problem,
it is customary to include thermal conduction that 
preferentially stabilizes small-scale perturbations 
(e.g., \citealt{pio04,pio05}), and also broadens the transition
layer between cold and warm phases (e.g., \citealt{beg90}). 
\citet{koy04} studied TI in an initially static medium
and showed that numerical results converge only if these conductive
interfaces  are well resolved.
They found this requires the cell size to be less than one third of 
the Field length.  
In modeling  systems with large turbulent 
(e.g., \citealt{gaz05}) or ordered (this work) velocity 
flows over the grid, numerical conductivity may be 
considerable if resolution is inadequate.
The convergence study presented in \S\ref{sec:mf} 
suggests that the convergence criterion of \citet{koy04} is
valid in a moving medium, too, with the condition that 
the grid spacing must be smaller than the {\it effective}
Field length based on the total (physical $+$ numerical) 
conductivity.

In earlier work,
\citet{tub80} and \citet{mar83} performed time-dependent simulations of
 galactic spiral shocks that included cooling and heating, 
but their models appear to suffer from large numerical
conductivity associated with insufficient resolution. 
While they showed that phase transitions from warm gas to cold clouds
do occur in some models at the shock locations, they were unable to 
resolve the post-shock transition zone.  In particular, the 
$n_0=0.5\pcc$ model shown in Figure 7 of \citet{mar83} exhibits 
a smooth density profile without any indication of TI although much of 
the region is occupied by gas with density and temperature 
in the unstable range.\footnote{The heating and cooling functions adopted by
\citet{mar83} have the critical densities demarcating the warm, 
intermediate-temperature,
and cold phases about 20 times smaller than those in the present work. 
Thus, their model with $n_0=0.5\pcc$ corresponds to our model SC10, which
develops multi-phase structure.} 
This is presumably because large numerical conductivity 
(due to low resolution) renders
the time and/or length scales of TI 
in their simulations 
longer than the duration and/or width of the post-shock transition zone.
Indeed, we find that when we run our own models at 
low resolution (e.g. with 512 zones or less for model 
SC10), the numerical conductivity is
large enough that 
the gas in the transition zone does not undergo TI, and instead smoothly
converts to interarm warm phase.
 
In terms of their time-averaged properties, the overall dynamics 
and flow characteristics of spiral shocks with TI remain similar 
to those of isothermal spiral shocks (e.g., \citealt{rob69,shu73,woo75,
kim02}).  The salient features of spiral shocks with thermal 
evolution  include 
the facts that they allow phase transitions at the shock front 
(from warm to cold)
and in the post-shock transition zone (from cold to warm), 
and that there are
cold clumps in the interarm region.  Given the
very high density and pressure within the arm, 
cold gas downstream from the shock front
would transform to molecular clouds if self-gravity and chemical reactions 
for molecule formation were included.
By running SPH simulations with separate 
cold and warm ``particles'', 
\citet{dob07} indeed found that the density of the cold component inside 
arms is sufficient to form molecules
(see also \citealt{dob08}).  These authors did not allow for phase 
transitions between the cold and warm components, however, which 
is an essential aspect of the evolution (cf. \citealt{shu72}).

A post-shock transition zone, where the gas is predominantly in
the intermediate-temperature range, is a necessary feature of any
quasi-steady state. 
About 40\% of the atomic 
gas in the local Milky Way is observed to be CNM
\citep{hei01,hei03}, which is consistent with our results for mean
density $n_0=1\pcc$, similar to the local Milky Way value.\footnote{The
value of $n_0$ that produces the numerical results consistent with 
observations may differ in models with different $L_x$ since the 
fractional size of the transition region
is likely to 
depend on the ratio of cooling time to arm-to-arm crossing time.}
\citet{hei03} also find that about half of the remaining atomic gas is
consistent with being thermally-stable WNM, while the balance is in
the unstable temperature range 500-5000K.  These fractions are also
comparable to what we find for $n_0=1\pcc$.  The presence of
thermally-unstable gas has been interpreted as owing to 
dynamical effects which act on timescales comparable to the heating
and cooling times.  Dynamical processes that have been investigated include
magnetorotational instability \citep{pio04,pio05,pio07}, colliding flows 
\citep{aud05,hei05,hei06,vaz06}, energy injection from OB stars
\citep{vaz00,gaz01}, and supernova explosions \citep{avi05,mac05}.
Here, we have shown that even without these additional small-scale 
energy sources, a significant amount of intermediate-temperature 
gas forms as a natural product of large-scale spiral shocks with TI,
primarily in the expanding region downstream from the arm.

Many numerical studies of TI have shown  that turbulent amplitudes driven
by ``pure TI'' {\it alone} are quite small.  For example, 
\citet{pio04} showed that pure TI in an initially static, uniform medium 
in thermal equilibrium 
produces only a modest level of random gas motions, 
$\sigma \sim  0.2-0.3\kms$.  
\citet{kri02a,kri02b} similarly found that TI of gas starting from
thermal equilibrium, or with time-dependent heating, results in 
only subsonic turbulence, $\sigma \sim0.2\kms$ when the cold gas dominates.
The velocity dispersions $\sigma\sim1.5\kms$ of the gas found 
in the interarm and transition zones in our one-dimensional models 
are significantly larger than those from pure TI.  
These velocities are much lower than those
we found in previous (isothermal) simulations \citep{kim06,kko06} 
in which the vertical shock structure is resolved, however.
Evidently, shock flapping  offers a more effective means of converting
galactic rotation to turbulence than simple in-plane oscillations. 
Work is currently underway to see how TI interacts with
vertically-resolved spiral shocks -- and differential buoyancy of cold
and warm gas -- to generate turbulence and structure
in the ISM.

\acknowledgments
We thank A.\ Kritsuk for drawing our attention to earlier work of
Marochnik et al. on spiral shocks with thermal instability.
This work was supported by the Korea Research Foundation Grant funded
by the Korean Government (MOEHRD) (KRF -- 2007 -- 313 -- C00328).  The work of 
E.~C.~O on this project was supported by the U.~S.\ National Science
Foundation under grant AST0507315.
The numerical computations presented in this work were performed on the
Linux cluster at KASI (Korea Astronomy and Space Science Institute)
built with funding from KASI and ARCSEC.

\clearpage

\begin{deluxetable}{lccclcclcclccclccc}
\rotate
\tabletypesize{\footnotesize}
\tablecaption{Summary of model parameters and simulation results\label{tbl:model}} \tablewidth{0pt}
\tablehead{
\colhead{Model} &
\colhead{$n_0$} &
\multicolumn{2}{c}{Arm} &
\colhead{} &
\multicolumn{2}{c}{Transition} &
\colhead{} &
\multicolumn{2}{c}{Interarm} &
\colhead{} &
\multicolumn{3}{c}{Mass Fractions (\%)} &
\colhead{} &
\multicolumn{3}{c}{Volume Fractions (\%)} \\
\cline{3-4}\cline{6-7}\cline{9-10}\cline{12-14}\cline{16-18}
\colhead{}&
\colhead{}&
\colhead{$\!\!\Delta L/L_x\!\!\!\!$} &
\colhead{$\!\!\Delta t/\tcro\!\!\!\!$} &
\colhead{}&
\colhead{$\!\!\Delta L/L_x\!\!\!\!$} &
\colhead{$\!\!\Delta t/\tcro\!\!\!\!$} &
\colhead{}&
\colhead{$\!\!\Delta L/L_x\!\!\!\!$} &
\colhead{$\!\!\Delta t/\tcro\!\!\!\!$} &
\colhead{}&
\colhead{Cold} &
\colhead{$\!\!\!\!$Intermediate$\!\!\!\!$} &
\colhead{Warm} &
\colhead{}&
\colhead{Cold} &
\colhead{$\!\!\!\!$Intermediate$\!\!\!\!$} &
\colhead{Warm} \\
\colhead{(1)}&
\colhead{(2)}&
\colhead{(3)} &
\colhead{(4)} &
\colhead{}&
\colhead{(5)} &
\colhead{(6)} &  
\colhead{}&
\colhead{(7)} &
\colhead{(8)} &
\colhead{}&
\colhead{(9)} &
\colhead{(10)} &
\colhead{(11)} &
\colhead{}&
\colhead{(12)} &
\colhead{(13)} &
\colhead{(14)} 
}
\startdata
  SW0.1 &    0.1 & \nodata & \nodata & & \nodata & \nodata & & \nodata & \nodata & &  0 &    0 &      100 &  &      0 &    0 &     100 \\
  SW0.5 &    0.5 &  0.00 & 0.05 & & 0.04 & 0.14 & & 0.96 & 0.82 & &    18 &     26 &     55 &  &      0 &     10 &     90 \\
    SW1 &    1.0 &  0.00 & 0.10 & & 0.10 & 0.21 & & 0.89 & 0.69 & &    43 &     25 &     31 &  &      2 &     16 &     83 \\
    SU2 &    2.0 &  0.01 & 0.14 & & 0.16 & 0.22 & & 0.83 & 0.64 & &    59 &     27 &     14 &  &      4 &     26 &     70 \\
    SU3 &    3.0 &  0.01 & 0.16 & & 0.22 & 0.24 & & 0.77 & 0.60 & &    62 &      30 &     8 &  &      7 &     35 &     58 \\
    SU4 &    4.0 &  0.02 & 0.19 & & 0.26 & 0.25 & & 0.72 & 0.56 & &    65 &      29 &     5 &  &     11 &     40 &     49 \\
    SU5 &    5.0 &  0.03 & 0.20 & & 0.28 & 0.25 & & 0.69 & 0.54 & &    66 &      30 &     3 &  &     14 &     43 &     43 \\
    SU6 &    6.0 &  0.04 & 0.22 & & 0.32 & 0.28 & & 0.63 & 0.50 & &    67 &      31&     2 &  &     18 &     48 &     34 \\
    SU7 &    7.0 &  0.05 & 0.23 & & 0.35 & 0.30 & & 0.60 & 0.47 & &    70 &      29&      2 &  &     23 &     47 &     30 \\
    SU8 &    8.0 &  0.06 & 0.24 & & 0.35 & 0.29 & & 0.59 & 0.47 & &    73 &      26 &     1 &  &     29 &     45 &     26 \\
   SC10 &   10.0 &  0.08 & 0.27 & & 0.40 & 0.32 & & 0.52 & 0.41 & &    76 &      24 &     1 &  &     37 &     46 &     17 \\
   SC20 &   20.0 & \nodata & \nodata & & \nodata & \nodata & & \nodata & \nodata & &   100 &      0 &      0 &  &    100 &      0 &      0 \\
\enddata
\tablecomments{Col.\ (1): Model name;
Col.\ (2): Initial number density in units of $\cm^{-3}$;
Cols.\ (3)-(8): Width and duration of the arm, transition, 
and interarm regions.
Cols.\ (9)-(14): Mass and volume fractions of the cold, 
intermediate-temperature, and warm phases
}
\end{deluxetable}

\end{document}